\documentclass[sigconf]{acmart}

\usepackage{algorithm}
\usepackage{algpseudocode}
\usepackage{graphicx} 
\usepackage{subcaption}

\AtBeginDocument{%
  \providecommand\BibTeX{{%
    \normalfont B\kern-0.5em{\scshape i\kern-0.25em b}\kern-0.8em\TeX}}}

\setcopyright{acmlicensed}
\copyrightyear{2018}
\acmYear{2018}
\acmDOI{XXXXXXX.XXXXXXX}

\acmConference[Conference acronym 'XX]{Make sure to enter the correct
  conference title from your rights confirmation emai}{June 03--05,
  2018}{Woodstock, NY}
\acmISBN{978-1-4503-XXXX-X/18/06}

\begin{document}

\title{SiamQuality: A ConvNet-Based Foundation Model for Imperfect Physiological Signals}

\author{Cheng Ding}

\affiliation{%
  \institution{Department of Biomedical Engineering, Georgia Institute of Technology}
  \city{Atlanta}
  \state{GA}
  \country{USA}
}
\email{chengding@gatech.edu}

\author{Zhicheng Guo}
\affiliation{%
  \institution{Department of Electrical and Computer Engineering, Duke University}
  \city{Durham}
  \state{NC}
  \country{USA}}

\author{Zhaoliang Chen}
\affiliation{%
  \institution{Department of Computer Science, Emory University}
  \city{Atlanta}
  \state{GA}
  \country{USA}
}

\author{Randall J Lee}
\affiliation{%
 \institution{School of Medicine, University of California at San Francisco}
 \city{San Francisco}
 \state{CA}
 \country{USA}
 }

\author{Cynthia Rudin}
\affiliation{%
  \institution{Department of Computer Science, Duke University}
\city{Durham}
  \state{NC}
  \country{USA}}

\author{Xiao Hu}
\affiliation{%
  \institution{Nell Hodgson Woodruff School of Nursing, Emory University}
  \city{Atlanta}
  \state{GA}
  \country{USA}
}
\email{xiao.hu@emory.edu}

\renewcommand{\shortauthors}{Ding et al.}

\begin{abstract}
Foundation models, especially those using transformers as backbones, have gained significant popularity, particularly in language and language-vision tasks. However, large foundation models are typically trained on high-quality data, which poses a significant challenge, given the prevalence of poor-quality real-world data. This challenge is more pronounced for developing foundation models for physiological data;  such data  are often noisy, incomplete, or inconsistent. The present work aims to provide a toolset for developing foundation models on physiological data. We leverage a large dataset of photoplethysmography (PPG) signals from hospitalized intensive care patients. For this data, we propose SimQuality, a novel self-supervised learning task based on convolutional neural networks (CNNs) as the backbone to enforce representations to be similar for good and poor quality signals that are from similar physiological states. 
We pre-trained the SimQuality on over 36 million 30-second PPG pairs and then fine-tuned and tested on six downstream tasks using external datasets. The results demonstrate the superiority of the proposed approach on all the downstream tasks, which are extremely important for heart monitoring on wearable devices.
Our method indicates that CNNs can be an effective backbone for foundation models that are robust to training data quality.

\end{abstract}




\received{20 February 2007}
\received[revised]{12 March 2009}
\received[accepted]{5 June 2009}

\maketitle

\section{Introduction}

Foundation models, particularly those with transformer architectures as the backbone \cite{transformer}, have significantly influenced the landscape of artificial intelligence in recent years. Their remarkable ability to capture and generate human language, as well as to process and interpret complex visual information, has set new standards in language and language-vision tasks \cite{llm_survey1, llm_survey2, llm_vision_survey1, llm_vision_survey2}. However, these large-scale models need to be trained with massive high-quality data, which are collected from books, journal articles, and internet sources. These text data typically follow rules of grammar and semantics, which allows the model's outputs to be coherent, contextually relevant, and linguistically accurate. 


However, other real world domains exhibit substantial amounts of poor-quality data. This gap between the ideal training data quality and the actual quality of available data makes it difficult to train and deploy foundation models. This challenge becomes even more daunting in the context of physiological data analysis \cite{physio_quality}. The reliability of physiological signals is often compromised by the presence of artifacts and noise, which can stem from patient movement, sensor displacement, and physiological variations. These artifacts can significantly distort the signal, leading to inaccurate assessments in clinical diagnostics and remote health monitoring. The inherent noise, incompleteness, and inconsistency, compounded by the complexities of human physiology and the unpredictable conditions of free-living data collection environments \cite{Charlton2023}, necessitate robust model architectures to address these data quality issues.

\begin{figure}[htbp]
\begin{center}
\includegraphics[width=8cm]{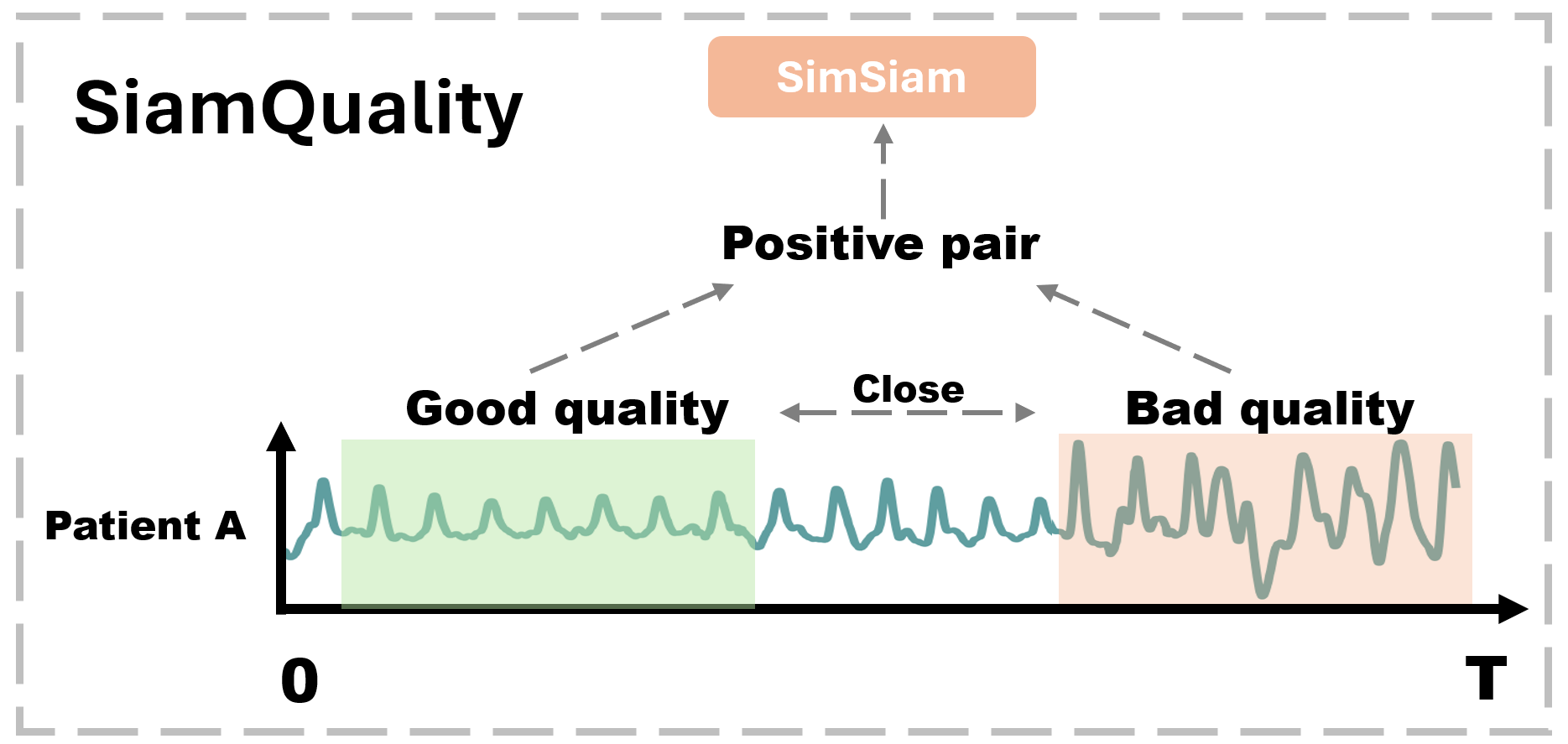}
\caption{The proposed SiamQuality to address the data quality issue in physiological data. }
\label{fig:fig1}
\Description{This methodology leverages the power of contrastive learning to extract robust features from PPG signals, irrespective of their quality. The rationale behind is, one poor quality PPG signal segments represents the same underlying physiological information with its good quality neighbors if they are collected within a limited temporal window from the same patient.}
\end{center}
\end{figure}

In this study, we propose SiamQuality, a new approach to address the issue of physiological data quality when developing foundation models. Instead of using transformers, which are complex and difficult to troubleshoot, we utilize CNNs as the backbone, given their demonstrated excellent ability to represent sequences \cite{cnn_ts1, cnn_ts2, cnn_ts3}. If the goal is solely representational learning, then CNNs provide an efficient framework to build foundational models. To train and test our approach, we leverage a large dataset of photoplethysmography (PPG) signals \cite{Charlton2023} collected from hospitalized intensive care patients. 

As shown in Figure \ref{fig:fig1}, the proposed SiamQuality is based on the SimSiam architecture \cite{Chen2021}, with a novel signal quality-based pairing mechanism. This methodology leverages the power of contrastive learning to extract robust features from PPG signals, alleviating the influence of poor signal quality. The rationale behind this approach is that \textbf{representations of PPG signals from similar physiological states, as learned by CNNs, should be similar} and that \textbf{a bad-quality PPG signal in a temporal vicinity of a good-quality one from the same human is likely to be from a similar physiological state}.

As a foundation model, SiamQuality meets three criteria: \textbf{i) Pre-trained on large-scale data}. In this study, we collected over 36 million 30-second PPG pairs (600,000+ hours) from 21,000 patients as the training data for SiamQuality. \textbf{ii) Accommodates Increasing Complexity}. We use CNN, specifically ResNet \cite{He2016}, as the backbone and successfully observe a performance improvement of SiamQuality when increasing model size. \textbf{iii) Adaptability}. We conducted extensive experiments on various tasks, including investigating different pre-training mechanisms and fine-tuning on six different PPG-based downstream classification and regression datasets. Experimental results validate that our method improves performance on all downstream tasks, achieving state-of-the-art (SOTA) performance across a range of tasks. Notably, on the tasks of respiration rate and atrial fibrillation detection, it outperforms state-of-the-art models by 43\% and 7\% respectively.

This study goes beyond merely presenting an efficient model for PPG signals; it provides an alternative solution that emphasizes the often-overlooked aspect of data quality in the development of foundation models. It sets a precedent for addressing data quality issues more broadly in AI and machine learning.

\section{Related Work}

\textbf{Foundation models for physiological data}. There are numerous foundation models in many areas, including language and computer vision, though only a few of them focus on physiological data. 
HeartBeit utilizes masked image modeling to develop a transformer model based on vision for analyzing electrocardiogram (ECG) waveforms. This pre-training approach enhances the accuracy and detailed explainability of the model's predictions. Another foundation model, the Biosignal Transformer Model (BIOT) \cite{biot}, learns embeddings for biosignals, especially for electroencephalogram (EEG) signals. BIOT enables effective knowledge transfer across different datasets and allows joint training on multiple sources. In the work \cite{frozen_ecg}, The authors enhance the ECG encoder by fusing the latent space generated from a frozen large language model applied to text from ECG reports. Consequently, their method does not rely on the specific type of annotated data and can be seamlessly transferred to any new database.

\textbf{Contrastive learning for time series data}. Positive and negative pairs are often needed in self-supervised contrastive learning. For instance, TS2Vec \cite{Yue2022} adopts a hierarchical contrastive learning framework to handle time series augmentations, where each time step is represented as a point in an embedding space. CoST \cite{Woo2022} encodes disentangled trend and seasonal representations through contrastive learning. The above-mentioned studies require negative pairs for the self-contrastive learning, where a negative pair consists of parts from two different time series. However, SimTS \cite{Zheng2023} have conducted experiments showing that current ways of constructing negative pairs are not efficient for time series data, meaning that the performance of the model does not benefit from the negative pairs. Instead, they propose that two consecutive time series can be combined as one positive pair, where positive pairs are pulled together in latent space. With SimSiam, which only requires positive pairs, SimTS has outperformed TS2Vec \cite{Yue2022} and CoST \cite{Woo2022} in experiments of \cite{Zheng2023}. In our study, we also use the same SimSiam structure and have included a modified SimTS \cite{Zheng2023} as one of the baselines, discussed in Subsection \ref{subsec:sim}.

\subsection{Method}
\subsection{Preliminaries and Problem Formulation}

 Our dataset $K= \{x_i,y_i,t_i\}_{i=1}^N$ consists of PPG strips $x_i$, each associated with a label $y_i$ representing the percentage of artifacts (perturbation in signal) in $x_i$ (ranging from 0\% to 100\%), where $t_i$ is the collection time. Our goal is to learn a robust representation that is minimally affected by artifacts.

\subsection{Signal Quality Pairing}
\begin{figure}[htbp]
\begin{center}
\includegraphics[width=8cm]{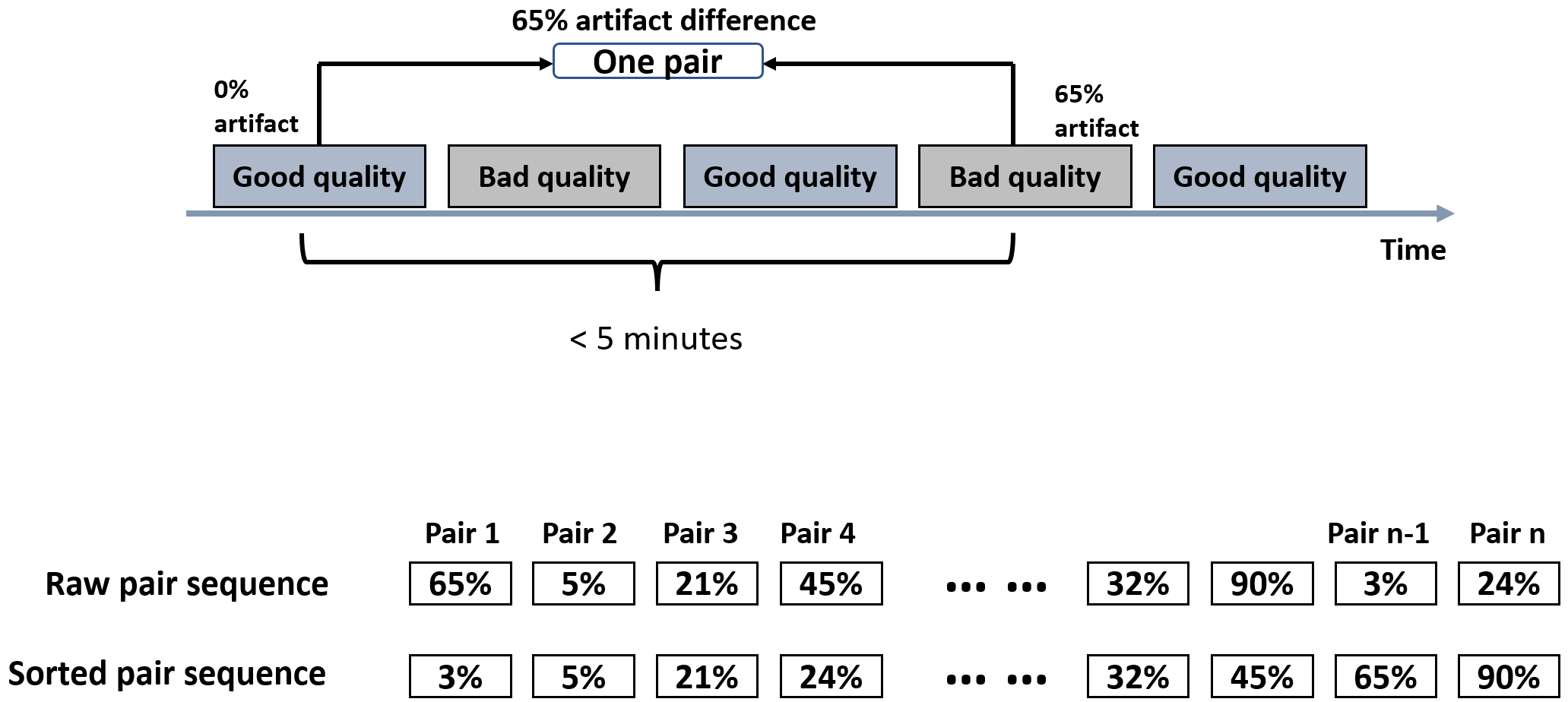}
\caption{The mechanism for signal quality pairing}
\label{fig2}
\Description{xxxxx}
\end{center}
\end{figure}

For our study, we utilized continuous PPG recordings collected from  adult intensive care units (ICUs), generating pairs of varying quality as illustrated in Figure \ref{fig2}. To categorize each 30-second PPG segment's quality, we implemented a quality assessment method previously developed \cite{Chen2023}. This method classifies each segment into a spectrum of quality from good to bad, based on the percentage of artifacts present. For $x_i$, the collection time $t_i$  for each strip is also recorded. In our methodology, segments identified as good quality ($y_i=0$) serve as anchor points for beginning a search for a nearby low-quality signal.

From each anchor point, we search for segments within a 5-minute window that exhibits bad quality (high $y_i$). Among these, we select the segment with the highest temporal distance from the anchor, ensuring a varied degree of comparison in terms of artifacts and time. This approach allows us to construct pairs of PPG segments that are physiologically similar but differ significantly in their quality, as dictated by the presence of artifacts. These pairs are then used to train our model, enabling it to learn robust representations that are minimally affected by artifacts.

In addition, we implemented \textbf{curriculum learning} \cite{Wang2021CurrLearning} by systematically feeding the model training pairs, starting with small artifacts, and later introducing larger artifacts.
The size of the artifact between a clean and not-clean signal is quantified by a measure $C(x_1,x_2 )=|y_1- y_2 |$. After sorting the training data based on C, the model is initially trained with pairs having lower $C(x_1,x_2)$ and progressively introduced to pairs with higher differences.

\subsection{Contrastive Learning Framework using SimSiam}
\begin{figure}[htbp]
\begin{center}
\includegraphics[width=7cm]{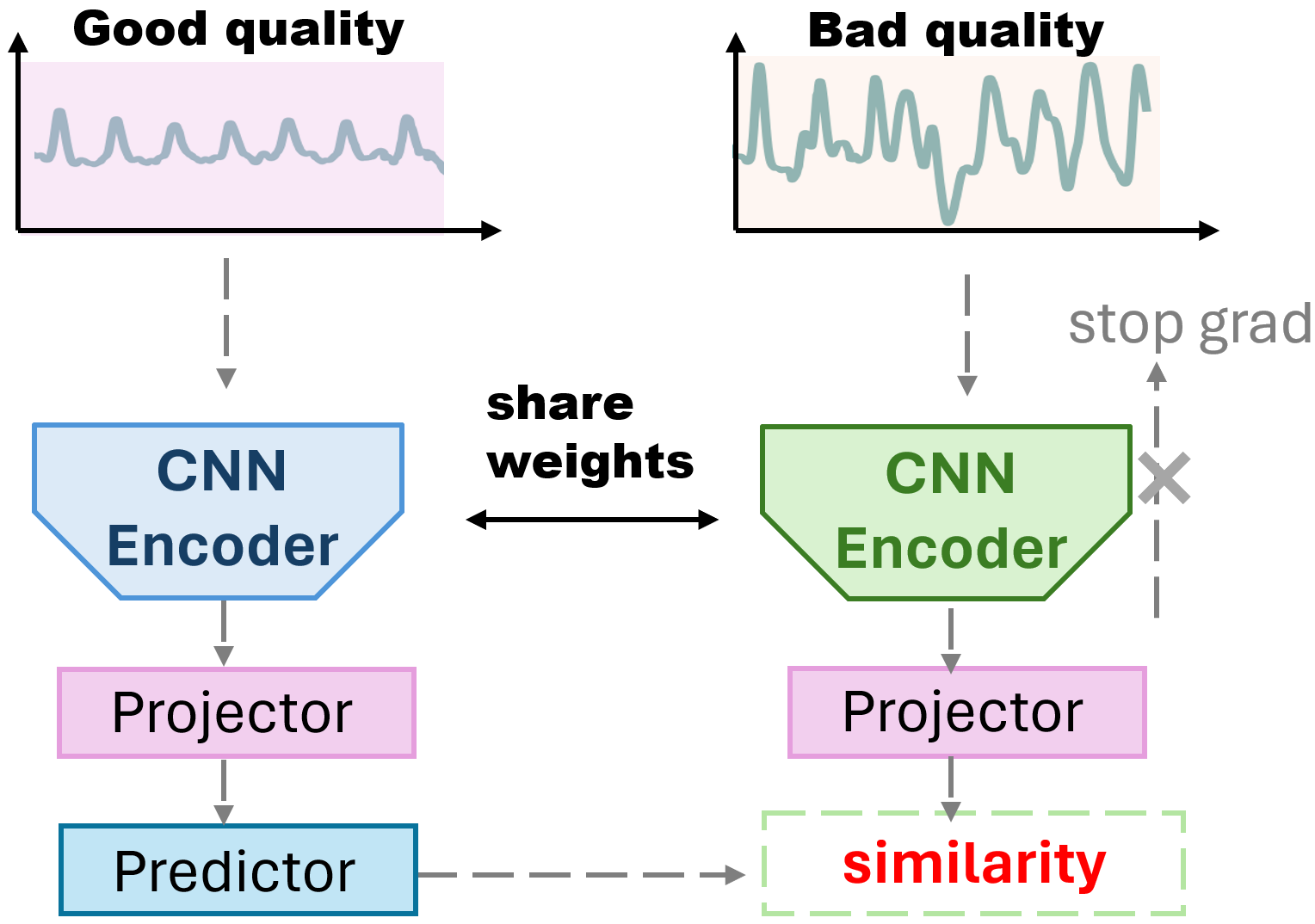}
\caption{Architecture for SimSiam with quality pairing augmentation}
\label{fig:fig3}
\Description{}
\end{center}
\end{figure}

Our approach is grounded in the SimSiam architecture \cite{Chen2021}, adapted for PPG signal processing, as shown in Figure \ref{fig:fig3}. The key components of our model include an encoder $E$, a projector $P$, and a predictor $D$.

\subsubsection{Encoder and Projector}
The encoder $E$ transforms the input signal  x into a representation $h=E(x)$. The projector P then maps this representation into a lower-dimensional space, $z=P(h)$, suitable for contrastive learning. 

\subsubsection{Predictor} The predictor D maps the projected representation z to the space where contrastive loss is applied. For a pair of signals $x_1$ (good quality) and $x_2$ (bad quality), their latent representations are $z_1=P(E(x_1 ))$ and $z_2=P(E(x_2 ))$, respectively. The predictor’s output for $x_i$ is $p_i=D(z_i)$.

\subsubsection{Contrastic Loss Function} The cosine similarity loss $L_{cosine}$ is used to minimize the distance between the predicted latent feature vectors within the pair:
$$L_{cosine}(x_1, x_2) = -\frac{p_1\cdot z_2}{\|p_1\|_2\cdot \|z_2\|_2} - \frac{p_2\cdot z_1}{\|p_2\|_2\cdot \|z_1\|_2}$$
where $\cdot$ denotes the dot product, and $\|\cdot\|_2$ represents the $L_2$ norm. 

\begin{algorithm}
\caption{SimSiam-Based PPG Signal Processing with Curriculum Learning}
\begin{algorithmic}[1] 
\State \textbf{Input:} Dataset \( K = \{x_i, y_i, t_i\}^N_{i=1} \) of PPG signals \( x_i \) with artifact percentage \( y_i \) and collection time \( t_i \)
\State \textbf{Output:} Specify the output if needed
\State \textbf{Step 1:} Initialization
\State Initialize Encoder \( E \), Projector \( P \), Predictor \( D \), and Classifier
\State Define cosine similarity loss function \( L_{cosine} \)
\State \textbf{Step 2:} Data Preparation
\For{\textbf{each} PPG signal \( (x_i, y_i = 0) \)}
    \State Find PPG \( x_j \) that 
    \begin{center}
        \( j = \max\{k | (|t_k - t_i| < 5 \text{mins}) \textrm{ AND } (y_k > 0.2)\} \)
    \end{center}
    \State Pair PPG signals \( (x_i, x_j, c_i) \) where \( c = y_j - y_i \)
\EndFor

\State \textbf{Step 3:} Curriculum Learning and Model Training
\For{\textbf{each} pair \( (x_{good}, x_{bad}) \)}
    \State Encode signals: \( h_{good} = E(x_{good}), h_{bad} = E(x_{bad}) \)
    \State Project encoded signals: \( z_{good} = P(h_{good}), z_{bad} = P(h_{bad}) \)
    \State Predictor: \( p_{good} = D(z_{good}), p_{bad} = D(z_{bad}) \)
    \State Compute contrastive loss: \( L = \frac{1}{2} (L_{cosine}(p_{good}, z_{bad}) + L_{cosine}(p_{bad}, z_{good})) \)
    \State Update model parameters to minimize \( L \)
\EndFor
\end{algorithmic}
\end{algorithm}

\section{Experiments and Results}
\subsection{Data}

The dataset used for model pre-training was obtained from routine patient monitoring systems in the intensive care units of the University of California, San Francisco (UCSF) Medical Center. This collection was conducted with an approved waiver of written patient consent under the UCSF Institutional Review Board (IRB number: 14-13262). The retrospective use of this fully de-identified data is conducted according to the terms of a data use agreement
between UCSF and Emory University.

The waveform dataset, which includes physiological signals, cardiac arrhythmia alarms, and linked electronic health records (EHR) from over 24,100 patients, was collected at the UCSF Medical Center between March 2013 and December 2018. The demographics of the patients in this dataset are reported in Table \ref{tab:participant_characteristics}. In the waveform dataset, a total of 2.6 million hours of continuous signals were collected, including seven-lead ECG (I, II, III, V, AVR, AVL, AVF), one-channel PPG signal, and one-channel respiration rate. All signals are sampled at 240Hz.

\begin{table}[h]
  \caption{Characteristics of Participants Enrolled in UCSF medical center}
  \label{tab:participant_characteristics}
  \begin{tabular}{lcc}
    \toprule
    Characteristics & Total Cohort (N = 28539) \\
    \midrule
    \textbf{Sex} & \\
    Female & 13203 (46.2\%) \\
    Male & 15330 (53.7\%) \\
    Others & 6 (0\%) \\
    \addlinespace
    \textbf{Age} & \\
    $\geq$65 yr & 12157 (42.6\%) \\
    55-64 yr & 5370 (18.8\%) \\
    40-54 yr & 4372 (25.3\%) \\
    22-39 yr & 2715 (9.5\%) \\
    $<$22 yr & 3925 (13.8\%) \\
    \addlinespace
    \textbf{Race or ethnic group} & \\
    White or Caucasian & 15890 (55.7\%) \\
    Black or African American & 2159 (7.4\%) \\
    Asian & 4364 (15\%) \\
    Unknown/Declined & 1149 (4\%) \\
    Others & 4913 (16.9\%) \\
    Native Hawaiian  & 426 (1.46\%) \\
    American Indian & 212 (0.7\%) \\
    \bottomrule
  \end{tabular}
\end{table}

\subsection{Data Preprocessing}

Continuous PPG recordings were divided into discrete, 30-second segments without overlap. Each of these segments was first downsampled to a frequency of 40Hz, followed by min-max normalization to standardize the signal range. To evaluate the quality of these signals, a previously developed binary PPG signal quality assessment tool \cite{pereira2019supervised} was employed, categorizing each segment as either `good quality' or `bad quality.' For segments classified as having bad quality, an additional analysis was applied using another signal quality segmentation model \cite{Chen2023}. This model is specifically designed to segment the locations of artifacts within each 30-second PPG signal, therefore it can be used to quantify the level of artifacts present in each bad quality signal.

\subsection{Downstream Tasks}
To test the effectiveness of our method, we selected a range of diverse public downstream tasks, allowing for reproducibility and validation of our results by the broader research community. Table \ref{tab:datasets} provides an overview of the datasets and their respective statistics.

\subsection{Performance Evaluation Metrics}
In our study, we employ distinct evaluation metrics tailored to the nature of the tasks. For regression tasks, the Mean Absolute Error (MAE) is used. For classification tasks, we utilize the F1 score, a balanced metric that considers both precision and recall, making it suitable for evaluating the performance of our classification models.

\begin{table}
  \caption{Statistics of public datasets for various downstream tasks}
  \label{tab:datasets}
  \begin{tabular}{lcc}
    \toprule
    Task & No. Subjects  \\
    \midrule
    \textbf{Heart Rate Estimation} & & \\
    \textbf{Signal Type: ECG, PPG, Accelerometer} &&\\
    TROIKA \cite{Zhang2014} & 12 \\
    Dalia \cite{Reiss2019} & 15  \\
    WESAD \cite{Schmidt2018} & 17 \\
    \addlinespace
    \textbf{Blood Pressure Estimation}& & \\
    \textbf{Signal Type:  ECG, PPG, ABP Waveforms} &&\\
    PulseDB (MIMIC-III+VitalDB) \cite{Wang2023} & 5361  \\
    \addlinespace
    \textbf{Respiration Rate Estimation}  & & \\
    \textbf{Signal Type: PPG} &&\\  
    BIDMC PPG and Respiration Dataset \cite{Pimentel2016} & 53\\
    \addlinespace
    \textbf{AF Detection} & & \\
    \textbf{Signal Type: PPG} &&\\
    Stanford AF dataset \cite{TorresSoto2020} & 148 \\
    \bottomrule
  \end{tabular}
\end{table}

\begin{figure}[ht]
\begin{center}
\includegraphics[width=6cm]{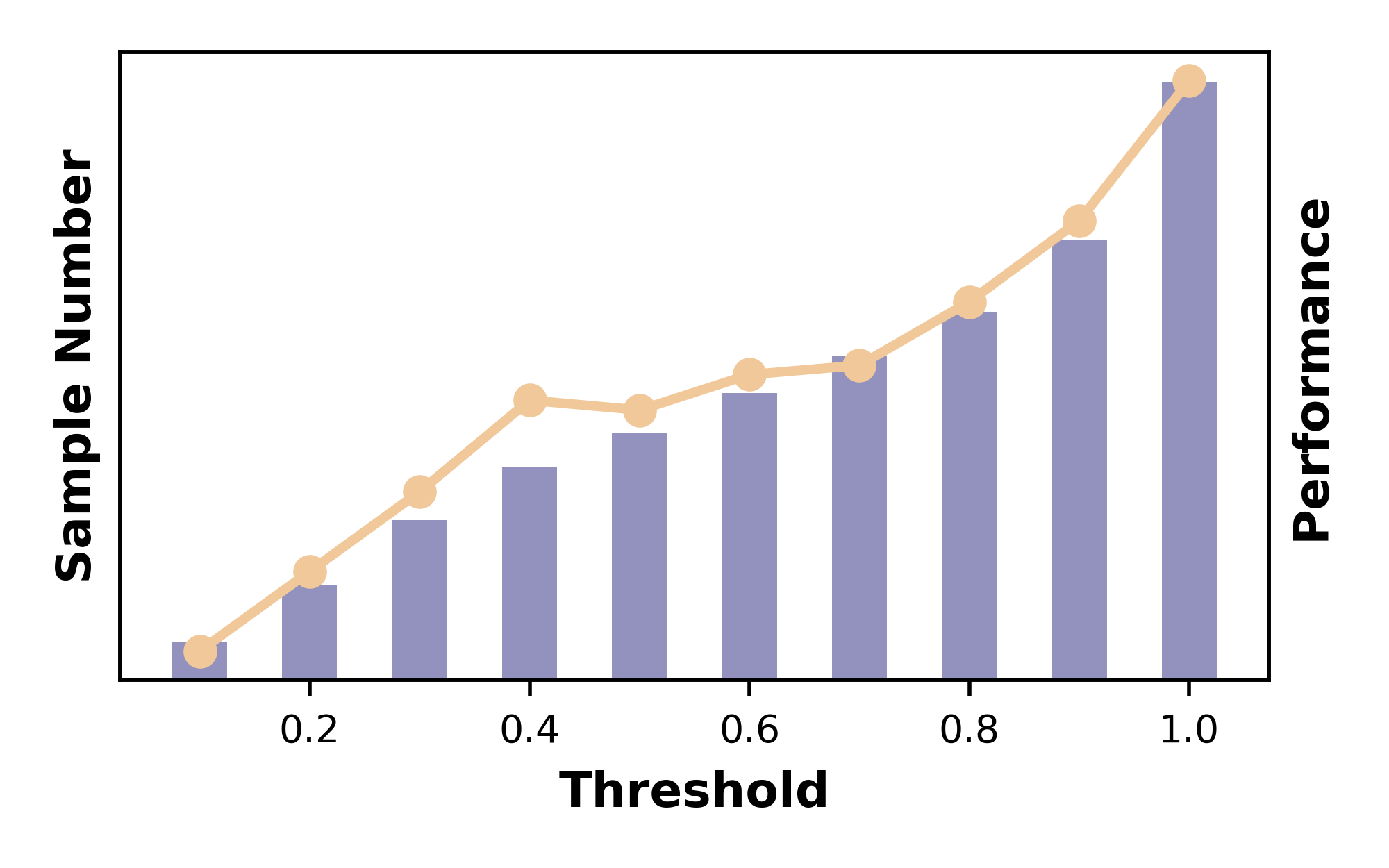}
\caption{AT-curve. The horizontal axis represents the upper limit of the signal quality for each subgroup. The height of each bar denotes the sample size for each respective subgroup. The line plot shows MAE within each subgroup.}
\label{fig:fig4}
\Description{xxxxx}
\end{center}
\end{figure}

Additionally, we introduce an innovative metric, the \textbf{Artifact Tolerance Curve} (AT-Curve), illustrated in Figure \ref{fig:fig4}. The AT-Curve is designed to assess the robustness of our models across varying levels of signal quality within the test data. The level of signal quality is calculated by the method of \citet{Chen2023}. The horizontal axis represents the upper limit for the signal quality of each subgroup, with values ranging from 0 to 1.0. A value of 1.0 represents the entire test set, while a value of 0 corresponds to the subset with the highest signal quality (i.e., the cleanest subset). The height of each bar on the AT-Curve denotes the sample size for each respective subgroup.
Overlaying the bars, a line plot traces the performance of our models—measured in terms of MAE for regression tasks and F1 score for classification tasks—within each of these subgroups. This line plot provides shows how our models' performance varies with the quality of the PPG signals. The AT curve thus serves a dual purpose: it not only illustrates the distribution of signal quality within our test data but also maps our models' performance against this distribution. The AT-curve offers a comprehensive view of our models' efficacy across a spectrum of signal qualities, highlighting their robustness and adaptability in real-world scenarios.

\subsection{Experimental Results}

\subsubsection{Comparison between Different Contrastive Learning Methods}
\label{subsec:sim}

\begin{table*}
  \caption{Comparison between different contrastive learning methods.}
  \label{tab:comparison_contrastive_learning}
  \begin{tabular}{lccccccc}
    \toprule
    Base Encoder: Resnet50 & Augmentation Method & \multicolumn{3}{c}{HR (MAE)} & RR (MAE) & AF (F1 Score) & BP (MAE) \\
    \cmidrule(r){3-5} \cmidrule(r){6-6} \cmidrule(r){7-7} \cmidrule(r){8-8}
     && TROIKA & Dalia & WESAD & BIDMC & Stanford & PulseDB \\
    \midrule
    SimCLR && 4.92 & 7.53 & 6.73 & 2.34 & 0.59 & 11.40 \\
    SwAV && 4.82 & 7.15 & 6.52 & 2.04 & 0.64 & 11.22 \\
    MoCo &Conventional& 4.70 & 7.28 & 6.53 & 2.17 & 0.63 & 11.17 \\
    BYOL &Augmentation& 4.97 & 7.50 & 6.77 & 2.41 & 0.61 & 11.40 \\
    SimSiam && 4.71 & 7.34 & 6.65 & 2.33 & 0.63 & 10.71 \\
    \addlinespace
    SimSiam &Temporal Sampling (Modified SimTS)& 4.92 & 7.31 & 6.54 & 2.08 & 0.63 & 10.94 \\
     &+ Quality pairing& 4.73 & 7.15 & 6.35 & 1.96 & 0.65 & 10.65 \\
     & + Curriculum learning&\textbf{4.69} & \textbf{7.02} & \textbf{6.27} & \textbf{1.84} & \textbf{0.68} & \textbf{10.32} \\
    \bottomrule
  \end{tabular}
\end{table*}

In the first experiment, we conducted a comprehensive evaluation of several self-supervised learning  (SSL)  models: SimCLR \cite{Chen2020}, SwAv \cite{Zhu2020}, MOCO \cite{He2020}, BYOL \cite{Grill2020}, and SimSiam \cite{Chen2021}. Methods in the top half of Table \ref{tab:comparison_contrastive_learning} all used the same augmentation method, which is to randomly select from (Gaussian noise, powerline noise, and flipping) within each mini-batch. Original signal and its augmented counterpart are combined as one positive pair, and signal with augmentation from another signal are combined as one negative pair.
Among the five SSL approaches, only BYOL and SimSiam do not require negative pairs. We chose SimSiam for the rest of the experiments that require only positive pairs since its performance exceeds that of BYOL. 

We now investigate different augmentation methods with SimSiam, as reported in the bottom half of Table \ref{tab:comparison_contrastive_learning}. Temporal sampling, instead of only using two consecutive signal as one pair (SimTS \cite{Zheng2023}), we modify the approach that uses one good quality PPG as an anchor point, and randomly selects another signal within 5 minutes, regardless of signal quality. Then, we added consideration of signal quality into the temporal sampling, and we further added curriculum learning as the third augmentation method. 

The results show a trend of improvement with each successive augmentation strategy, for all datasets, for all of the tasks.  \textbf{SimSiam with quality-pairing and curriculum learning consistently yielded the best results,} for all of the tasks (heart rate prediction, RR prediction, AF detection, and BP prediction).

\subsubsection{SimSiam with Different Model Sizes}

We investigated the scaling laws of SimQuality, as reported in Table \ref{tab:simsiam-model-size}. The results reveal a clear trend: as the complexity of the ResNet architecture \cite{He2016} increases, there is a marked improvement in performance. This trend is evident across all metrics, including MAE for Heart Rate, Respiratory Rate, and Blood Pressure, as well as Atrial Fibrillation detection. ResNet152, being the most complex model, consistently exhibited the best performance. However, ResNet152 is more expensive to train than ResNet50. The experiments in the remaining subsections do not require as much computation as earlier experiments, so we use ResNet152 from now on.


\begin{table}
  \caption{SimSiam with different model sizes}
  \label{tab:simsiam-model-size}
  \centering
  \begin{tabular}{lccc}
    \toprule
    Dataset & Resnet50 & Resnet101 & Resnet152 \\
    \midrule
    TROIKA (MAE) & 4.69 & 4.66 & \textbf{4.59} \\
    Dalia (MAE) & 7.02 & 6.92 & \textbf{6.80} \\
    WESAD (MAE) & 6.27 & 6.07 & \textbf{5.88} \\
    BIDMC (MAE) & 1.84 & 1.53 & \textbf{0.89} \\
    Stanford (F1) & 0.68 & 0.69 & \textbf{0.71} \\
    PulseDB (MAE) & 10.32 & 9.64 & \textbf{8.60} \\
    \bottomrule
  \end{tabular}
\end{table}

\begin{table}
  \caption{Results compared with SOTA in HR datasets (MAE)}
  \label{tab:hrdatasets}
  \centering
  \begin{tabular}{lccc}
    \toprule
    Datasets & TROIKA & Dalia & WESAD \\
    \midrule
    Deep PPG \cite{Reiss2019} & 4.00 & 7.65 & 7.47 \\
    \textcolor{gray}{PPGNet*} \cite{Shyam2019} & 3.36 & - & - \\
    \textcolor{gray}{CurToSS*} \cite{Zhou2020} & 2.20 & 5.00 & 6.40 \\
    \textcolor{gray}{TAPIR*} \cite{Huang2020} & 2.50 & 4.60 & 4.20 \\
    \textcolor{gray}{BeliefPPG*} \cite{Bieri2023} & \textbf{1.88} & \textbf{4.26} & \textbf{3.81} \\
    Our methods & 4.59 & 6.80 & 5.88 \\
    \bottomrule
  \end{tabular}
\end{table}

\subsubsection{Comparing with previous work on each downstream task}
We compared our model's performance against the previous SOTA results for each specific task. Our model leverages the SimSiam framework, augmented with a ResNet152 encoder, and employs Quality-Pairing with Curriculum Learning. We report performance metrics and  the AT-curve for each task. 

\begin{figure*}
    \centering
    \begin{subfigure}[b]{0.3\textwidth}
        \includegraphics[width=\textwidth]{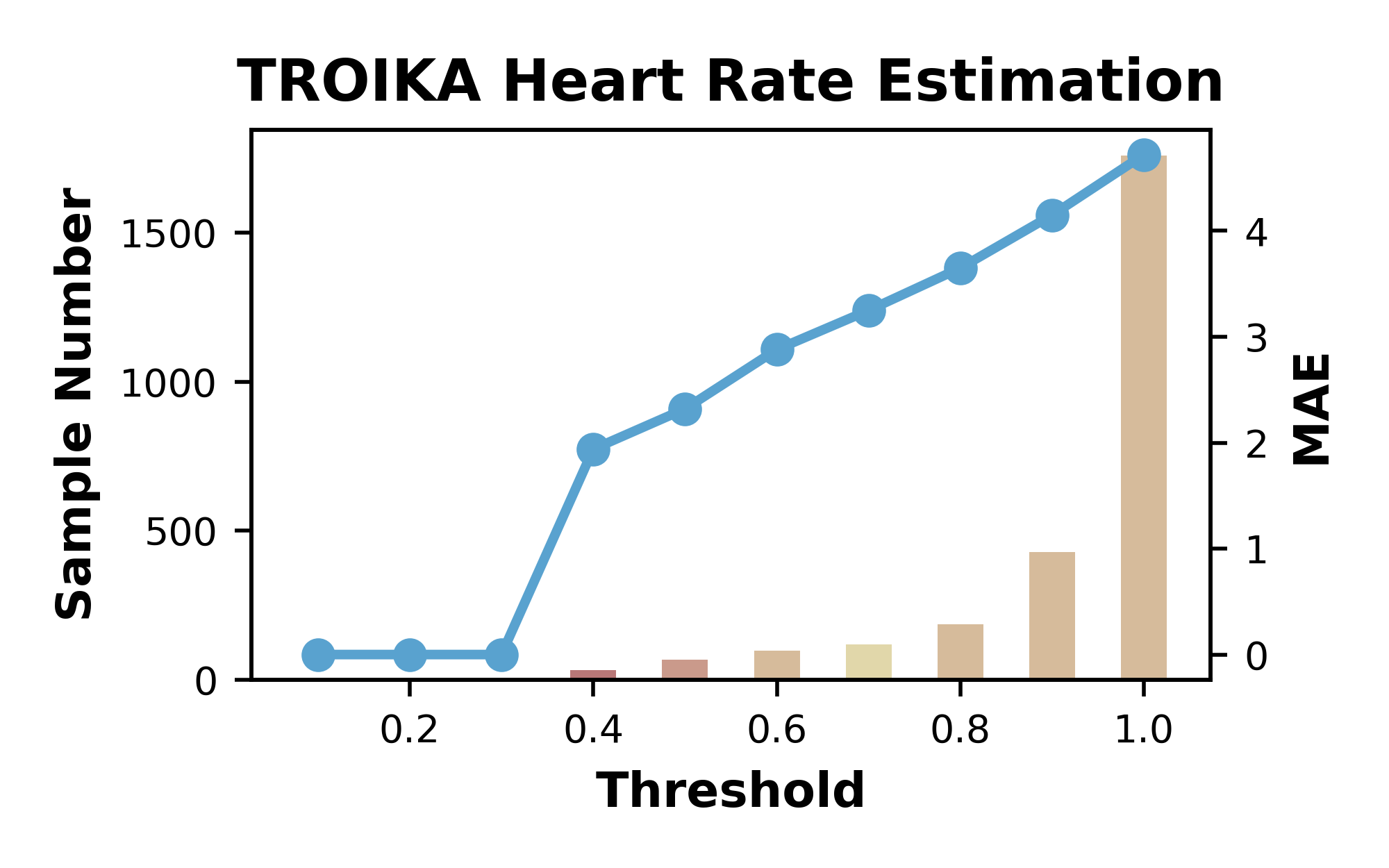}
        \caption{TROIKA Heart Rate} 
        \label{fig:sub1}
    \end{subfigure}
    \hfill 
    \begin{subfigure}[b]{0.3\textwidth}
        \includegraphics[width=\textwidth]{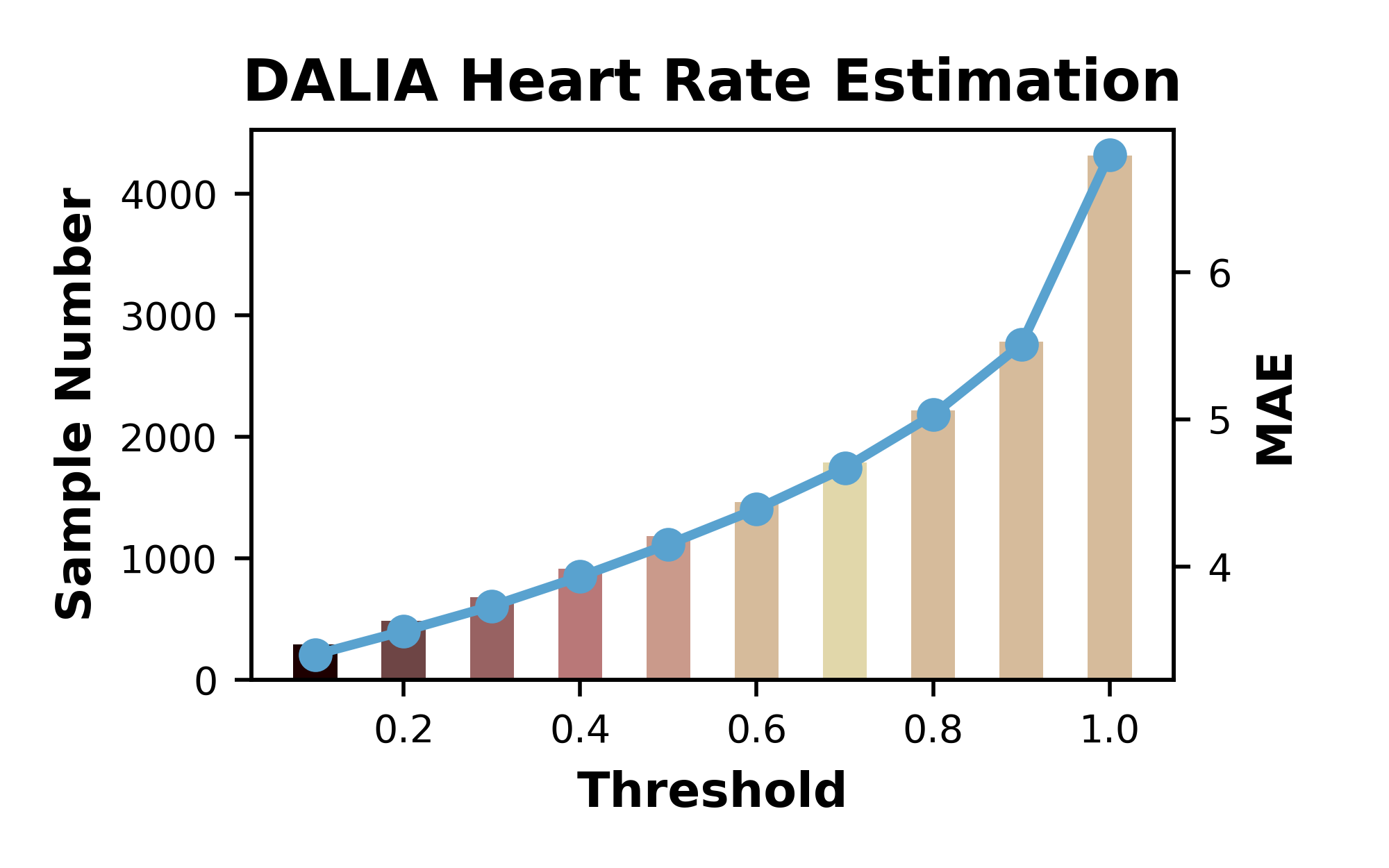}
        \caption{DALIA Heart Rate} 
        \label{fig:sub2}
    \end{subfigure}
    \hfill 
    \begin{subfigure}[b]{0.3\textwidth}
        \includegraphics[width=\textwidth]{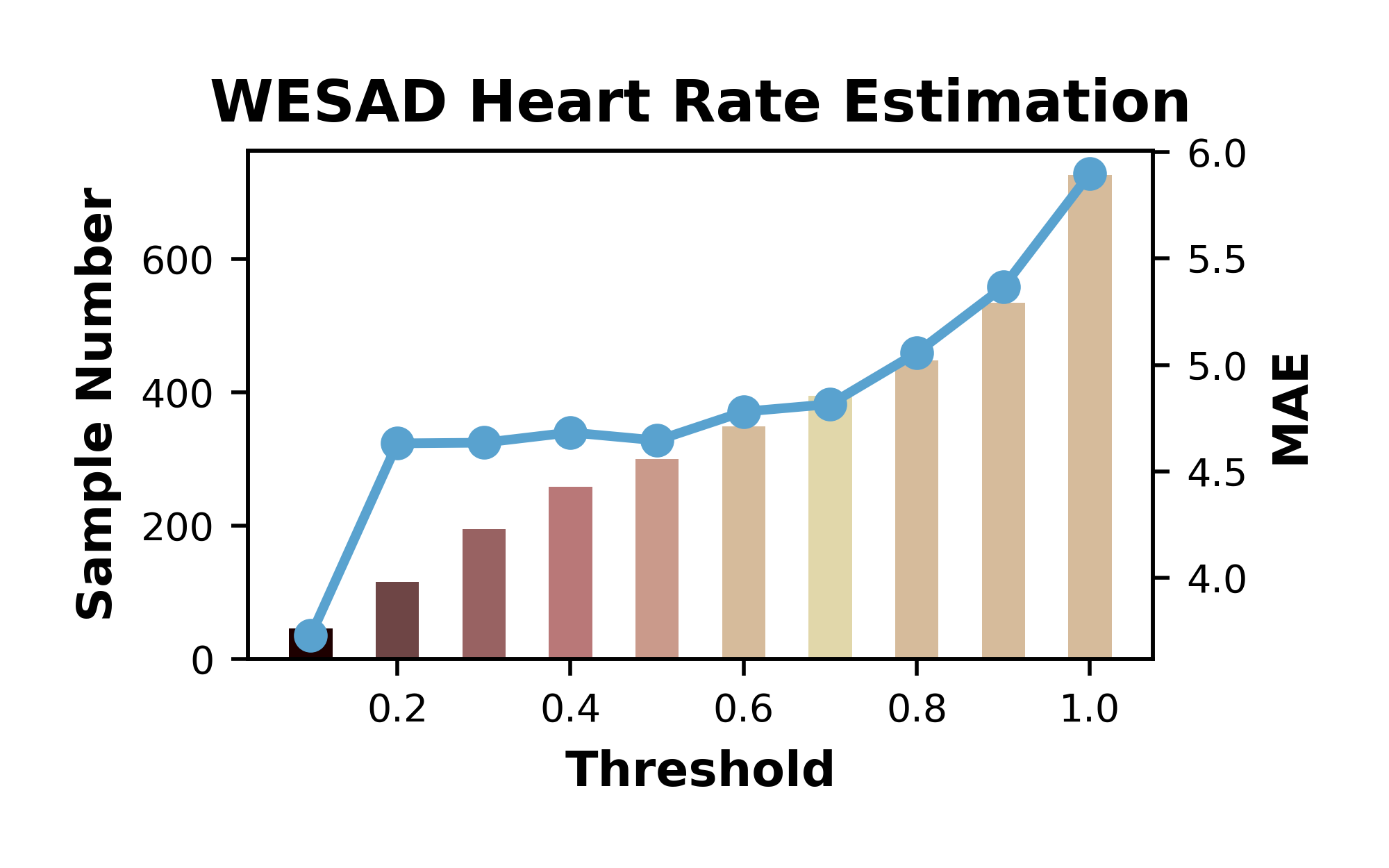}
        \caption{WESAD Heart Rate} 
        \label{fig:sub3}
    \end{subfigure}
 
    \begin{subfigure}[b]{0.3\textwidth}
        \includegraphics[width=\textwidth]{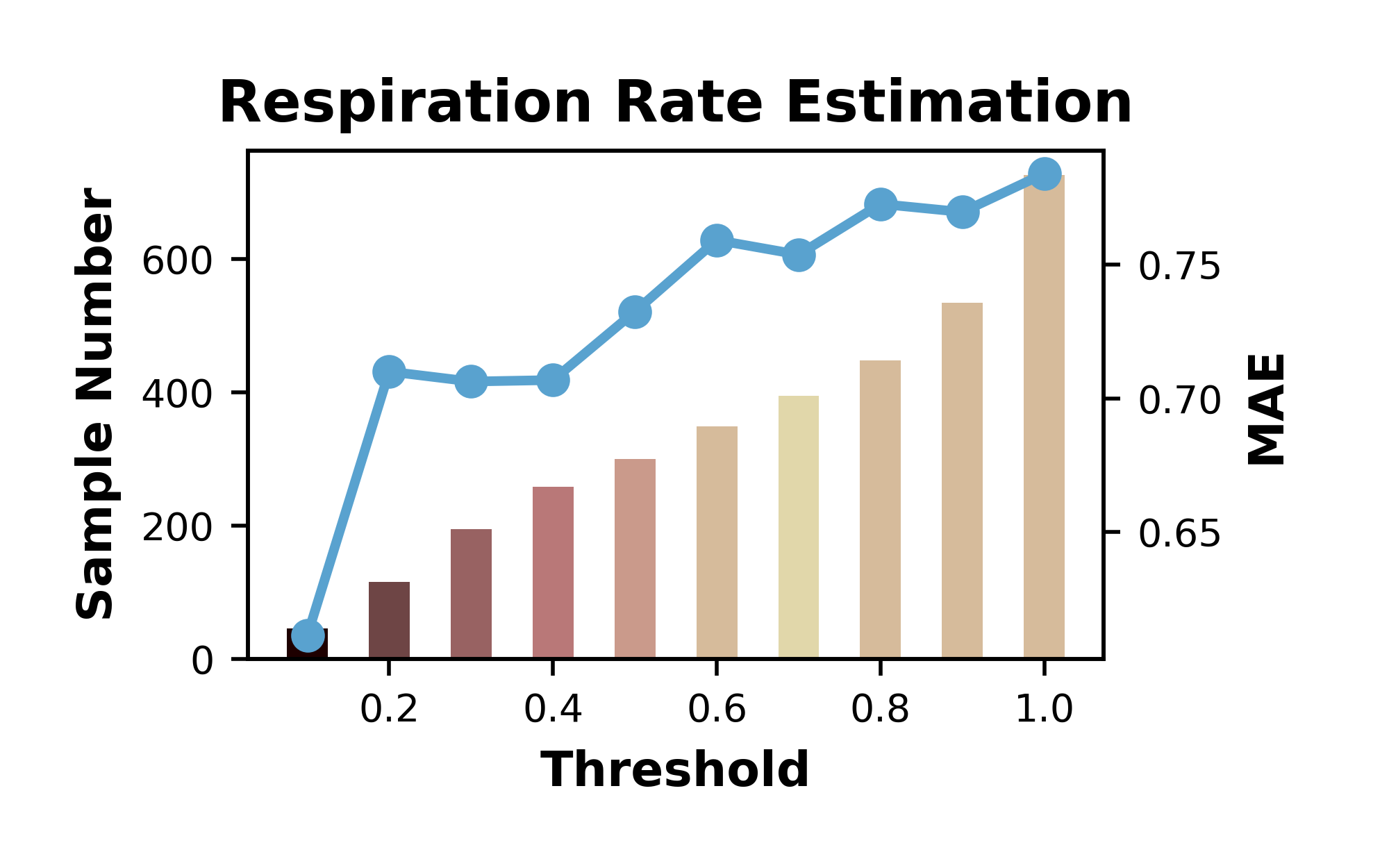}
        \caption{BIDMC Resipiration Rate} 
        \label{fig:sub4}
    \end{subfigure}
    \hfill
    \begin{subfigure}[b]{0.3\textwidth}
        \includegraphics[width=\textwidth]{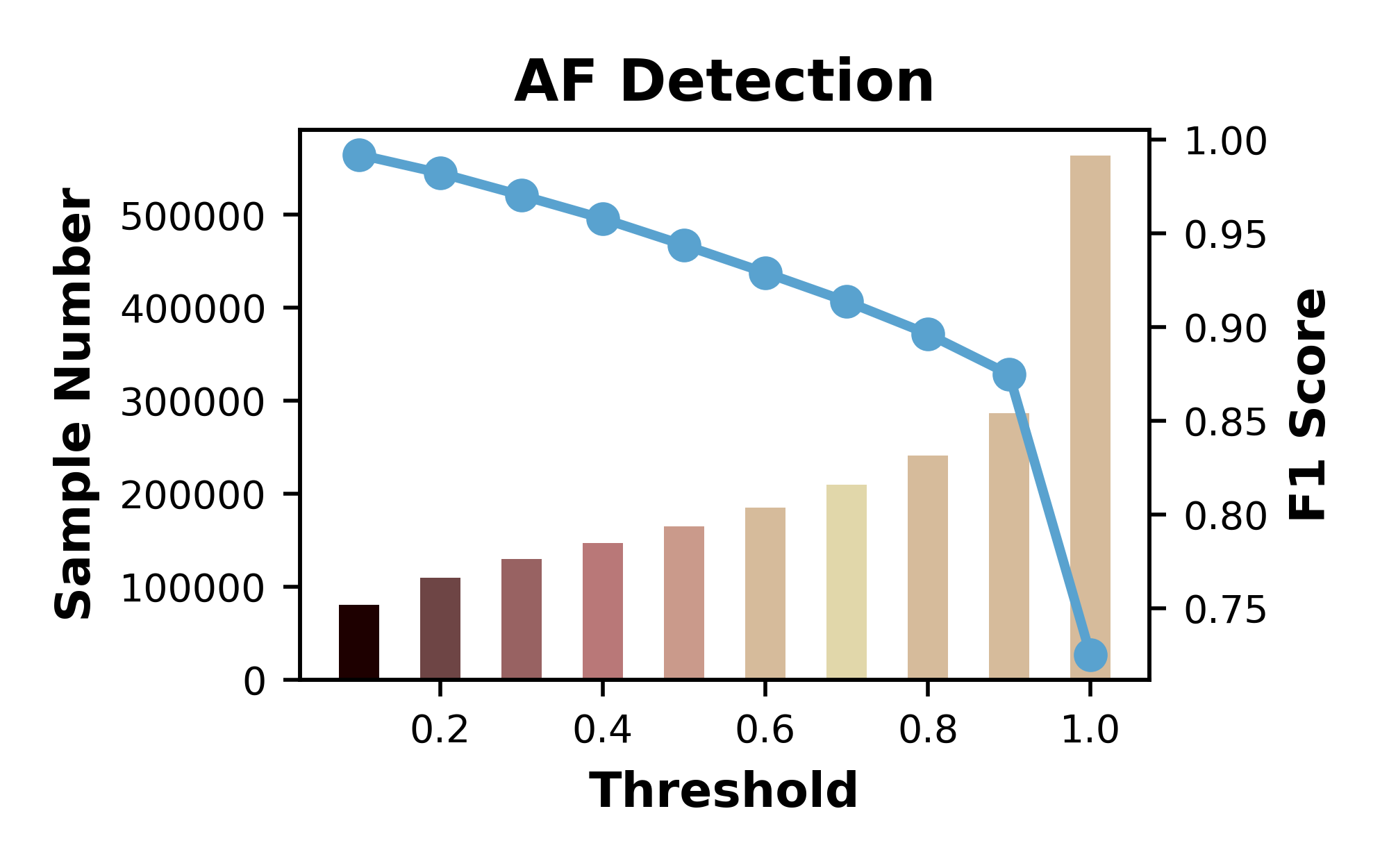}
        \caption{Stanford AF} 
        \label{fig:sub5}
    \end{subfigure}
    \hfill
    \begin{subfigure}[b]{0.3\textwidth}
        \includegraphics[width=\textwidth]{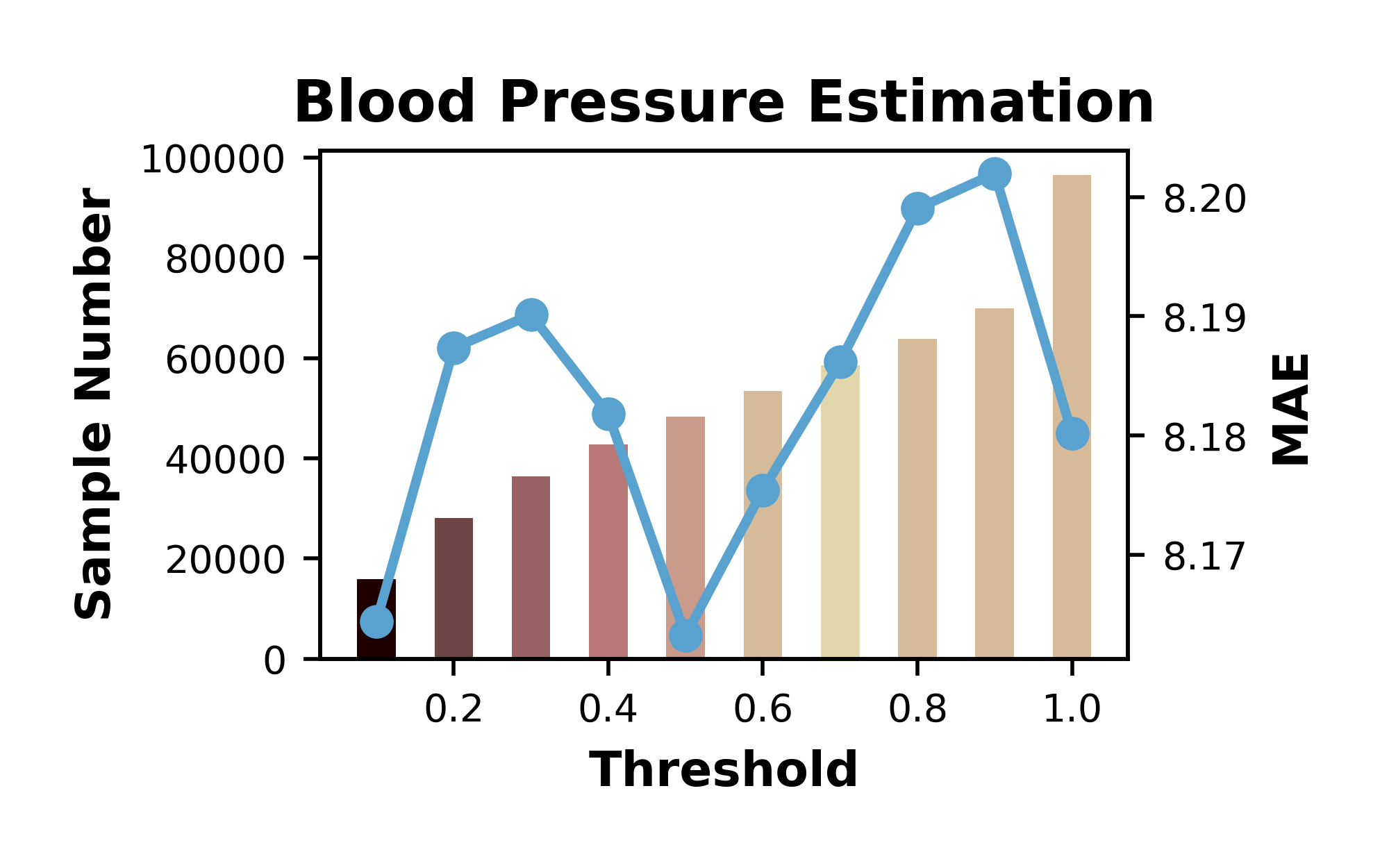}
        \caption{PulseDB Blood Pressure} 
        \label{fig:sub6}
    \end{subfigure}
    \caption{AT-Curve for all downstream tasks} 
    \label{fig:main}
\end{figure*}

\begin{table}
  \caption{Results compared with SOTA in RR dataset}
  \label{tab:rrdataset}
  \centering
  \begin{tabular}{lc}
    \toprule
    Methods & RR (MAE) \\
    \midrule
    Multiple auto-regression \cite{Pimentel2016} & 4.0 \\
    EEMD + PCA \cite{Nilsson2000} & 3.5 \\
    EEMD + Kalman filter \cite{Motin2017} & 1.9 \\
    Smart fusion \cite{Karlen2013} & 5.8 \\
    LSTM \cite{Kumar2022} & 1.51 \\
    LSTM+Attn \cite{Kumar2022} & 1.56 \\
    Our methods & \textbf{0.89} \\
    \bottomrule
  \end{tabular}
\end{table}

\begin{table*}[h]
  \caption{Results compared with SOTA in AF dataset}
  \label{tab:afdataset}
  \centering
  \begin{tabular}{lcc}
    \toprule
    Methods & AF detection (F1 Score) & Artifact Coverage \\
    \midrule
    Shen et al. 2019 \cite{Shen2019} & 0.684 & 100 \\
    Torres-Soto and Ashley (Deepbeat) \cite{TorresSoto2020} & 0.652 & 100 \\
    Torres-Soto and Ashley (Deepbeat, Non-poor) \cite{TorresSoto2020} & 0.646 & 27.67 \\
    Torres-Soto and Ashley (Deepbeat, Excellent only) \cite{TorresSoto2020} & 0.58 & 18.77 \\
    BayesBeat \cite{Das2022} & 0.671 & 100 \\
    BayesBeat (uncertainty threshold = 0.05) \cite{Das2022} & 0.754 & 54 \\
    Our methods & 0.71 & 100 \\
    \bottomrule
  \end{tabular}
\end{table*}

\begin{table}[h]
  \caption{Results compared with SOTA in SBP dataset}
  \label{tab:sbpdataset}
  \centering
  \begin{tabular}{lc}
    \toprule
    Method & SBP (MAE) \\
    \midrule
    CNN-GRU\textsubscript{attn} (ECG, PPG) \cite{Wang2023EMBC} & 4.90 \\
    UTransBPRNet (ECG, PPG) \cite{Zheng2023UTransBPNet} & \textbf{4.38} \\
    Our methods (PPG only) & 8.6 \\
    \bottomrule
  \end{tabular}
\end{table}

\begin{table*}[h]
  \caption{Results compared within different fine-tuning configurations}
  \label{tab:fine-tuning}
  \centering
  \begin{tabular}{lcccccc}
    \toprule
    Fine Tuning configuration & TROIKA & DALIA & WESAD & BP & AF & RR \\
    & (MAE) & (MAE) & (MAE) & (MAE) & (F1 Score) & (MAE) \\
    \midrule
    Fine-tune All & 4.6 & 6.8 & 5.8 & 8.6 & 0.71 & 0.89 \\
    Fine-tune Last & 5.78 & 8.55 & 7.3 & 10.5 & 0.61 & 2.2 \\
    & (+25.65\%) & (+25.74\%) & (+25.68\%) & (+22.09\%) & (-14.08\%) & (+147.19\%) \\
    In-domain pre-training & 4.84 & 7.15 & 6.1 & 9.3 & 0.64 & 1.6 \\
    + Fine-tune Last 
    & (+5.22\%) & (+5.11\%) & (+5.17\%) & (+8.14\%) & (-9.86\%) & (+79.78\%) \\
    \bottomrule
  \end{tabular}
\end{table*}

As reported in Table \ref{tab:hrdatasets}, we observe that discarding poor quality signals significantly impacts performance (* indicates poor-quality signals were discarded). Note that it is important to state that poor-quality signals were omitted because MAE values are not comparable between cases when they are omitted and when they are not. Comparing DeepPPG and our methods, which did not discard any signals, we perform better in Dalia and WESAD but are slightly worse in TROIKA. Through the AT curve shown in Figures 5a – 5c, we can see that DALIA and WESAD have a more uniform distribution of artifact levels within each bin, while TROIKA displays a significant skew towards noisy signals, with a majority of signals exhibiting over 90\% artifacts. In addition, from the AT-curve, we can observe that, in the case of DALIA, discarding signals containing over 60\% of artifacts allows us to surpass the performance of BeliefPPG. For WESAD, the threshold to exceed BeliefPPG involves discarding signals with artifact levels above 20\%. To outperform BeliefPPG in the TROIKA dataset, it is necessary to exclude signals with more than 40\% artifacts.

A range of methods are compared for RR prediction, as reported in Table \ref{tab:rrdataset}. Among all the previous state-of-the-art (SOTA) methods, the LSTM model demonstrated strong performance with an MAE of 1.51. However, our method outperformed all these techniques, achieving the lowest MAE of 0.89, indicating its superior accuracy in estimating Respiratory Rate. The AT curve presented in Figure 5d shows that performance does not uniformly improve with better signal quality. 

In the evaluation of AF detection methods, as reported in Table \ref{tab:afdataset}, additional artifact coverage for each method is provided in \cite{Das2022}. BayesBeat, with an F1 Score of 0.671 at full artifact coverage (without discarding any signals), significantly improved to 0.754 when applying an uncertainty threshold, albeit with reduced coverage (i.e., keeping only 54\% of the signals in the test set). Our methods outperformed many others, achieving an F1 Score of 0.71 with full artifact coverage, indicating a robust and accurate approach for AF detection across diverse signal conditions. Through the AT curve in Figure 5e, we can observe that over half of the signals have more than 90\% artifacts. The F1 score can reach up to 90\% if signals with over 80\% of artifacts are discarded.
 

In the study of PPG-based SBP estimation tasks, as reported in Table \ref{tab:sbpdataset}, the CNN-GRUattn method, which employs a blend of Convolutional Neural Networks and Gated Recurrent Units with attention mechanisms and uses both ECG and PPG signals, achieved a Mean Absolute Error (MAE) of 4.90. UTransBPNet, also integrating ECG and PPG data, improved upon this accuracy with an MAE of 4.38. In contrast, our method, which solely relies on PPG signals, recorded a higher MAE of 8.6, indicating a lower accuracy in SBP prediction compared to the multi-signal approaches. Even across the AT curve, the MAEs remain in a relatively narrow range, from 8.15 to 8.2. Hence, ECG adds value to SBP estimation. However, PPG is collected passively and can offer continuous monitoring, while ECG requires subjects to remain stationary and thus is not nearly as widely available.


\section{Pre-training and fine tune configuration}

For all the experiments discussed thus far, we fine-tuned all the parameters for each downstream task (we call this `fine-tune All'). We also considered a `fine-tune Last' strategy, where the parameters of the pre-trained encoder are frozen while only the newly initialized classifier layer is tuned according to the labels of target tasks).
We also tried a third strategy, where we pre-trained SiamQuality on the downstream datasets and then fine-tuned the classification layer.  (Note again that the pre-training does not require labels from downstream tasks, since the goal of placing bad- and good-quality signals requires only simulated labels -- it is a self-supervised method.)

As reported in Table \ref{tab:fine-tuning}, fine-tuning all layers of the model yields the best results across multiple metrics, including MAE in TROIKA, DALIA, WESAD, and BP predictions, as well as the F1 Score for AF detection. In contrast, fine-tuning only the last layer significantly diminishes the model's accuracy, increasing the MAE by about 25\% in most datasets and reducing the F1 Score for AF detection by 14.08\%. The hybrid approach of in-domain pre-training followed by fine-tuning the last layer offers some improvement over solely fine-tuning the last layer, but it still falls short of the effectiveness achieved by fine-tuning all layers.

\section{Visualization}
\begin{figure}
\begin{center}
\includegraphics[width=8cm]{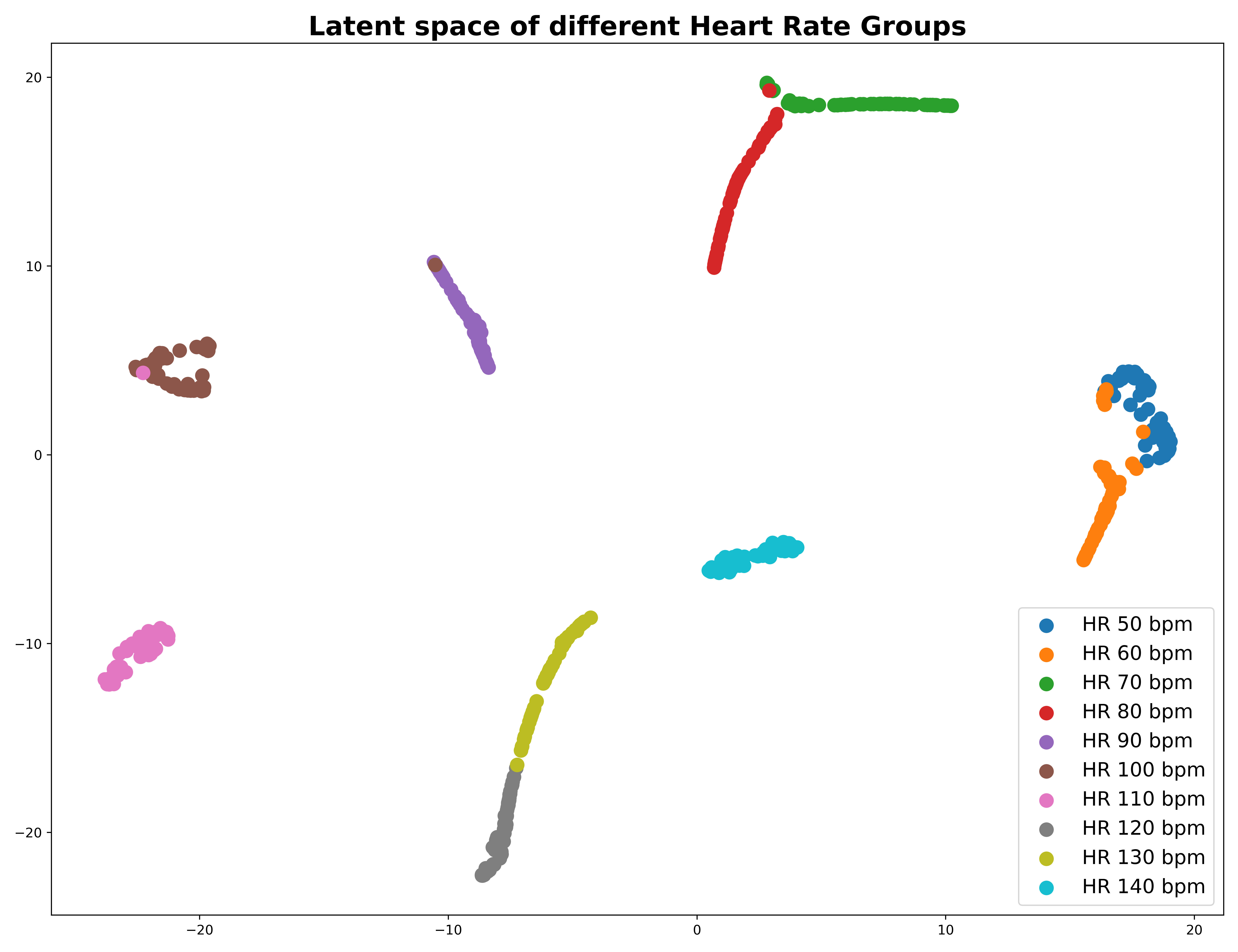}
\caption{Latent space for simulated PPG with various level of artifacts}
\label{fig:fig6}
\Description{}
\end{center}
\end{figure}

We explored the latent space of the SiamQuality model by generating simulated PPG signals with varying heart rates using NeuroKit2 \cite{Makowski2021}. To assess the model's robustness to noise, we incrementally introduced different levels of drift noise, motion amplitude noise, and powerline noise to the PPG signals, ranging from 0 (clean) to 0.7 (very noisy). After subjecting these simulated signals to the pre-trained model, we analyzed the resulting latent space. For dimension reduction and visualization, we employed PaCMAP \cite{Wang2021}, with the findings illustrated in Figure \ref{fig:fig6}. Notably, the results reveal that PPG signals with the same heart rate, despite varying noise levels, remain clustered together in the latent space. This observation underscores the efficacy of our proposed quality-pairing mechanism in mitigating the impact of artifacts in PPG signals on downstream physiological measurement tasks.

\section{Discussion}

In this study, we propose SiamQuality, a CNN-based foundation model for PPG signals. SiamQuality utilizes the SimSiam architecture but with a novel pairing mechanism, where we pair one good quality signal with its bad quality temporal neighbor. Curriculum learning was applied, where artifact level difference was increased over iterations to make the problem more difficult. After pre-training on 36 million PPG pairs, the model was then fine-tuned on six downstream datasets. Through extensive experiments, we demonstrate the capabilities of SiamQuality in terms of its contrastive learning architecture, scaling ability, and adaptability. The analysis of the results is discussed as follows:

First, our approach avoids selecting negative pairs (for example, pairs of signals from different individuals) to train the SimSiam architecture. This is because identifying helpful negative pairs for PPG model training is challenging \cite{Zheng2023}. As PPG signals exhibit strong periodicity, leading to the recurrence of similar patterns within and across patients, conventional augmentation methods may result in choosing inappropriate negative pairs. The same conclusion can be drawn from Table 3: conventional augmentation for time-series data, such as flipping and adding Gaussian noise, did not perform as well as temporal sampling and was further outperformed by quality-based pairing. This conclusion is in line with the results of \citet{Zheng2023}, who show that using conventional augmentation methods to construct negative pairs is not sufficient for time series data, and will lead to performance degradation.

Second, CNNs can serve as an effective backbone for foundation models that are robust to the quality of training data. Through our experiments, CNNs have demonstrated notable efficiency in processing and learning from data with inherent imperfections. This efficiency is attributed to their structural advantage in capturing local patterns through convolutional filters, which inherently makes them less sensitive to global noise and artifacts present in physiological signals.

Third, our previously developed artifact measurement tool \cite{Chen2023} plays a crucial role in this study by providing a quantifiable metric of data quality, thereby enabling curriculum learning and the generation of the AT-curve. Curriculum learning enables models to be better equipped to distill underlying patterns and relationships in the data, even when presented with signals heavily marred by noise and artifacts. The AT-curve allows a broader illustration of the model’s performance on physiological datasets with various artifact distributions.

As an early-stage investigation into foundation models for physiological data, our work has several limitations.

\textbf{Zero-shot learning is not available in our setting}. The reason is that SiamQuality focuses on aligning the latent space between good and bad quality signals, rather than aiming for specific tasks, such as detecting AF or estimating RR. This means that while SiamQuality excels at generalizing across different qualities of data and ensuring robustness to artifacts and noise, it does not inherently possess the ability to recognize or classify conditions it has not been explicitly trained on.

\textbf{Class imbalance problem}. We encountered a failed case in PPG-based Premature Ventricular Contraction (PVC) detection where the data provided in \cite{ppg_pvc} was small and highly imbalanced (42 30-second PVC-PPG vs. 200 30-second normal-PPG segments). Our approach achieved a 52\% F1 Score, while the original study achieved over 95\%. Our model's struggle with PVC detection highlights the necessity for strategies that can mitigate the effects of class imbalance, such as synthetic data generation, oversampling of minority classes, or the implementation of cost-sensitive learning together with the fine-tuning stage. Addressing this issue is critical for ensuring that foundation models for physiological data can accurately and reliably identify a wide range of conditions, including those that are rare but clinically significant.

\textbf{The quality assessment tool is not universal}. The tool is specifically designed for PPG signals, so it requires tailored quality assessment tools for each type of physiological data, such as ECG and EEG. To make our work more generalizable and applicable to a broader range of physiological data, it is important to develop a universal quality assessment tool that can accurately identify and quantify the quality of signals in their respective domains.

\section{Deployment}
Our model has been deployed at the following website: \url{https://www.nursingdatascience.emory.edu/modelmeetsdata}. The demo of ModelMeetsData (M2D) can be found at \url{https://www.youtube.com/watch?v=yA4Ky7hFdfw}. In M2D, users can test various PPG based models we developed by uploading their own PPG signals, and M2D will provide the prediction (classification) or estimation (regression) from the selected model. Any data uploaded by users will not be retained by M2D after the session. Transition from theory to practical deployment on medical wearable devices involves significant additional steps. This includes advanced engineering, extensive testing and clinical trials, and collaboration with a major manufacturer, which are beyond the scope of this paper. 

\section{Reproductivity}
Due to the nature of the private health data involved, the training dataset is available upon request to the corresponding author. The six downstream datasets are publicly available. Additionally, the code associated with this project is made publicly available in \url{https://github.com/chengding0713/SiamQuality}.

\section{Conclusion}

SiamQuality addresses the often-neglected issue of data quality in foundational models. Our focus is on photoplethysmography (PPG) signals — a type of physiological data where quality concerns are particularly pronounced. SiamQuality is based on the SimSiam architecture by implementing a unique pairing mechanism that connects a high-quality signal with a closely following lower-quality counterpart. After pre-trained on a large-scale dataset with 36 million PPG signal pairs, SiamQuality is further fine-tuned on six downstream datasets. The outcomes of comprehensive testing highlight effectiveness of  contrastive learning, performance scalability with model size, and its exceptional performance across a range of applications. This study goes beyond merely presenting an efficient model for PPG signals; it can be used for many applications that involve low-quality data, setting a precedent for addressing data quality issues more broadly in AI and machine learning.

\section{Acknowledgement}
This work is partially supported by NIH grant award R01HL166233.

\bibliographystyle{ACM-Reference-Format}
\bibliography{manuscript-bib}


\begin{thebibliography}{45}


\ifx \showCODEN    \undefined \def \showCODEN     #1{\unskip}     \fi
\ifx \showDOI      \undefined \def \showDOI       #1{#1}\fi
\ifx \showISBNx    \undefined \def \showISBNx     #1{\unskip}     \fi
\ifx \showISBNxiii \undefined \def \showISBNxiii  #1{\unskip}     \fi
\ifx \showISSN     \undefined \def \showISSN      #1{\unskip}     \fi
\ifx \showLCCN     \undefined \def \showLCCN      #1{\unskip}     \fi
\ifx \shownote     \undefined \def \shownote      #1{#1}          \fi
\ifx \showarticletitle \undefined \def \showarticletitle #1{#1}   \fi
\ifx \showURL      \undefined \def \showURL       {\relax}        \fi
\providecommand\bibfield[2]{#2}
\providecommand\bibinfo[2]{#2}
\providecommand\natexlab[1]{#1}
\providecommand\showeprint[2][]{arXiv:#2}

\bibitem[A et~al\mbox{.}(2019)]%
        {Shyam2019}
\bibfield{author}{\bibinfo{person}{Shyam A}, \bibinfo{person}{Vignesh Ravichandran}, \bibinfo{person}{Preejith S.P}, \bibinfo{person}{Jayaraj Joseph}, {and} \bibinfo{person}{Mohanasankar Sivaprakasam}.} \bibinfo{year}{2019}\natexlab{}.
\newblock \showarticletitle{PPGnet: Deep network for device independent heart rate estimation from photoplethysmogram}. In \bibinfo{booktitle}{\emph{Annual International Conference of the IEEE Engineering in Medicine and Biology Society (EMBC)}}. IEEE, \bibinfo{pages}{1899--1902}.
\newblock


\bibitem[Bieri et~al\mbox{.}(2023)]%
        {Bieri2023}
\bibfield{author}{\bibinfo{person}{Valentin Bieri}, \bibinfo{person}{Paul Streli}, \bibinfo{person}{Berken~Utku Demirel}, {and} \bibinfo{person}{Christian Holz}.} \bibinfo{year}{2023}\natexlab{}.
\newblock \bibinfo{title}{BeliefPPG: Uncertainty-aware Heart Rate Estimation from PPG signals via Belief Propagation}.
\newblock \bibinfo{howpublished}{arXiv preprint arXiv:2306.07730}.
\newblock


\bibitem[Charlton et~al\mbox{.}(2023)]%
        {Charlton2023}
\bibfield{author}{\bibinfo{person}{Peter~H. Charlton}, \bibinfo{person}{John Allen}, \bibinfo{person}{Raquel Bailón}, \bibinfo{person}{Stephanie Baker}, \bibinfo{person}{Joachim~A. Behar}, \bibinfo{person}{Fei Chen}, \bibinfo{person}{Gari~D. Clifford}, \bibinfo{person}{David~A. Clifton}, \bibinfo{person}{Harry~J. Davies}, \bibinfo{person}{Cheng Ding}, {and} \bibinfo{person}{Xiaorong Ding}.} \bibinfo{year}{2023}\natexlab{}.
\newblock \showarticletitle{The 2023 wearable photoplethysmography roadmap}.
\newblock \bibinfo{journal}{\emph{Physiological measurement}} \bibinfo{volume}{44}, \bibinfo{number}{11} (\bibinfo{year}{2023}), \bibinfo{pages}{111001}.
\newblock


\bibitem[Chen et~al\mbox{.}(2023b)]%
        {llm_vision_survey2}
\bibfield{author}{\bibinfo{person}{Fei-Long Chen}, \bibinfo{person}{Du-Zhen Zhang}, \bibinfo{person}{Ming-Lun Han}, \bibinfo{person}{Xiu-Yi Chen}, \bibinfo{person}{Jing Shi}, \bibinfo{person}{Shuang Xu}, {and} \bibinfo{person}{Bo Xu}.} \bibinfo{year}{2023}\natexlab{b}.
\newblock \showarticletitle{Vlp: A survey on vision-language pre-training}.
\newblock \bibinfo{journal}{\emph{Machine Intelligence Research}} \bibinfo{volume}{20}, \bibinfo{number}{1} (\bibinfo{year}{2023}), \bibinfo{pages}{38--56}.
\newblock


\bibitem[Chen et~al\mbox{.}(2023a)]%
        {Chen2023}
\bibfield{author}{\bibinfo{person}{Sully~F. Chen}, \bibinfo{person}{Zhicheng Guo}, \bibinfo{person}{Cheng Ding}, \bibinfo{person}{Xiao Hu}, {and} \bibinfo{person}{Cynthia Rudin}.} \bibinfo{year}{2023}\natexlab{a}.
\newblock \bibinfo{title}{Learned Kernels for Interpretable and Efficient PPG Signal Quality Assessment and Artifact Segmentation}.
\newblock \bibinfo{howpublished}{arXiv preprint arXiv:2307.05385}.
\newblock


\bibitem[Chen et~al\mbox{.}(2020)]%
        {Chen2020}
\bibfield{author}{\bibinfo{person}{Ting Chen}, \bibinfo{person}{Simon Kornblith}, \bibinfo{person}{Mohammad Norouzi}, {and} \bibinfo{person}{Geoffrey Hinton}.} \bibinfo{year}{2020}\natexlab{}.
\newblock \showarticletitle{A simple framework for contrastive learning of visual representations}. In \bibinfo{booktitle}{\emph{International Conference on Machine Learning}}. PMLR, \bibinfo{pages}{1597--1607}.
\newblock


\bibitem[Chen and He(2021)]%
        {Chen2021}
\bibfield{author}{\bibinfo{person}{Xinlei Chen} {and} \bibinfo{person}{Kaiming He}.} \bibinfo{year}{2021}\natexlab{}.
\newblock \showarticletitle{Exploring simple siamese representation learning}. In \bibinfo{booktitle}{\emph{Proceedings of the IEEE/CVF Conference on Computer Vision and Pattern Recognition}}. \bibinfo{pages}{15750--15758}.
\newblock


\bibitem[Clifford et~al\mbox{.}(2012)]%
        {physio_quality}
\bibfield{author}{\bibinfo{person}{GD Clifford}, \bibinfo{person}{J Behar}, \bibinfo{person}{Q Li}, {and} \bibinfo{person}{Iead Rezek}.} \bibinfo{year}{2012}\natexlab{}.
\newblock \showarticletitle{Signal quality indices and data fusion for determining clinical acceptability of electrocardiograms}.
\newblock \bibinfo{journal}{\emph{Physiological measurement}} \bibinfo{volume}{33}, \bibinfo{number}{9} (\bibinfo{year}{2012}), \bibinfo{pages}{1419}.
\newblock


\bibitem[Das et~al\mbox{.}(2022)]%
        {Das2022}
\bibfield{author}{\bibinfo{person}{Sarkar Snigdha~Sarathi Das}, \bibinfo{person}{Subangkar~Karmaker Shanto}, \bibinfo{person}{Masum Rahman}, \bibinfo{person}{Md.~Saiful Islam}, \bibinfo{person}{Atif Rahman}, \bibinfo{person}{Mohammad~Mehedy Masud}, {and} \bibinfo{person}{Mohammed~Eunus Ali}.} \bibinfo{year}{2022}\natexlab{}.
\newblock \showarticletitle{BayesBeat: Reliable atrial fibrillation detection from noisy photoplethysmography data}.
\newblock \bibinfo{journal}{\emph{Proceedings of the ACM on Interactive, Mobile, Wearable and Ubiquitous Technologies}} \bibinfo{volume}{6}, \bibinfo{number}{1} (\bibinfo{year}{2022}), \bibinfo{pages}{1--21}.
\newblock


\bibitem[Du et~al\mbox{.}(2022)]%
        {llm_vision_survey1}
\bibfield{author}{\bibinfo{person}{Yifan Du}, \bibinfo{person}{Zikang Liu}, \bibinfo{person}{Junyi Li}, {and} \bibinfo{person}{Wayne~Xin Zhao}.} \bibinfo{year}{2022}\natexlab{}.
\newblock \showarticletitle{A survey of vision-language pre-trained models}.
\newblock \bibinfo{journal}{\emph{arXiv preprint arXiv:2202.10936}} (\bibinfo{year}{2022}).
\newblock


\bibitem[Grill et~al\mbox{.}(2020)]%
        {Grill2020}
\bibfield{author}{\bibinfo{person}{Jean-Bastien Grill}, \bibinfo{person}{Florian Strub}, \bibinfo{person}{Florent Altché}, \bibinfo{person}{Corentin Tallec}, \bibinfo{person}{Pierre~H. Richemond}, \bibinfo{person}{Elena Buchatskaya}, \bibinfo{person}{Carl Doersch}, \bibinfo{person}{Bernardo~Avila Pires}, \bibinfo{person}{Zhaohan~Daniel Guo}, \bibinfo{person}{Mohammad~Gheshlaghi Azar}, \bibinfo{person}{Bilal Piot}, \bibinfo{person}{Koray Kavukcuoglu}, \bibinfo{person}{Rémi Munos}, {and} \bibinfo{person}{Michal Valko}.} \bibinfo{year}{2020}\natexlab{}.
\newblock \showarticletitle{Bootstrap your own latent-a new approach to self-supervised learning}. In \bibinfo{booktitle}{\emph{Advances in Neural Information Processing Systems}}, Vol.~\bibinfo{volume}{33}. \bibinfo{pages}{21271--21284}.
\newblock


\bibitem[Han et~al\mbox{.}(2020)]%
        {ppg_pvc}
\bibfield{author}{\bibinfo{person}{Dong Han}, \bibinfo{person}{Syed~Khairul Bashar}, \bibinfo{person}{Fahimeh Mohagheghian}, \bibinfo{person}{Eric Ding}, \bibinfo{person}{Cody Whitcomb}, \bibinfo{person}{David~D McManus}, {and} \bibinfo{person}{Ki~H Chon}.} \bibinfo{year}{2020}\natexlab{}.
\newblock \showarticletitle{Premature atrial and ventricular contraction detection using photoplethysmographic data from a smartwatch}.
\newblock \bibinfo{journal}{\emph{Sensors}} \bibinfo{volume}{20}, \bibinfo{number}{19} (\bibinfo{year}{2020}), \bibinfo{pages}{5683}.
\newblock


\bibitem[He et~al\mbox{.}(2020)]%
        {He2020}
\bibfield{author}{\bibinfo{person}{Kaiming He}, \bibinfo{person}{Haoqi Fan}, \bibinfo{person}{Yuxin Wu}, \bibinfo{person}{Saining Xie}, {and} \bibinfo{person}{Ross Girshick}.} \bibinfo{year}{2020}\natexlab{}.
\newblock \showarticletitle{Momentum contrast for unsupervised visual representation learning}. In \bibinfo{booktitle}{\emph{Proceedings of the IEEE/CVF Conference on Computer Vision and Pattern Recognition}}. \bibinfo{pages}{9729--9738}.
\newblock


\bibitem[He et~al\mbox{.}(2016)]%
        {He2016}
\bibfield{author}{\bibinfo{person}{Kaiming He}, \bibinfo{person}{Xiangyu Zhang}, \bibinfo{person}{Shaoqing Ren}, {and} \bibinfo{person}{Jian Sun}.} \bibinfo{year}{2016}\natexlab{}.
\newblock \showarticletitle{Deep residual learning for image recognition}. In \bibinfo{booktitle}{\emph{Proceedings of the IEEE Conference on Computer Vision and Pattern Recognition}}. \bibinfo{pages}{770--778}.
\newblock


\bibitem[Huang and Selvaraj(2020)]%
        {Huang2020}
\bibfield{author}{\bibinfo{person}{Nicholas Huang} {and} \bibinfo{person}{Nandakumar Selvaraj}.} \bibinfo{year}{2020}\natexlab{}.
\newblock \showarticletitle{Robust ppg-based ambulatory heart rate tracking algorithm}. In \bibinfo{booktitle}{\emph{Annual International Conference of the IEEE Engineering in Medicine \& Biology Society (EMBC)}}. IEEE, \bibinfo{pages}{5929--5934}.
\newblock


\bibitem[Karlen et~al\mbox{.}(2013)]%
        {Karlen2013}
\bibfield{author}{\bibinfo{person}{Walter Karlen}, \bibinfo{person}{Srinivas Raman}, \bibinfo{person}{J~Mark Ansermino}, {and} \bibinfo{person}{Guy~A Dumont}.} \bibinfo{year}{2013}\natexlab{}.
\newblock \showarticletitle{Multiparameter respiratory rate estimation from the photoplethysmogram}.
\newblock \bibinfo{journal}{\emph{IEEE Transactions on Biomedical Engineering}} \bibinfo{volume}{60}, \bibinfo{number}{7} (\bibinfo{year}{2013}), \bibinfo{pages}{1946--1953}.
\newblock


\bibitem[Kumar et~al\mbox{.}(2022)]%
        {Kumar2022}
\bibfield{author}{\bibinfo{person}{Amit~Krishan Kumar}, \bibinfo{person}{M. Ritam}, \bibinfo{person}{Lina Han}, \bibinfo{person}{Shuli Guo}, {and} \bibinfo{person}{Rohitash Chandra}.} \bibinfo{year}{2022}\natexlab{}.
\newblock \showarticletitle{Deep learning for predicting respiratory rate from biosignals}.
\newblock \bibinfo{journal}{\emph{Computers in Biology and Medicine}}  \bibinfo{volume}{144} (\bibinfo{year}{2022}), \bibinfo{pages}{105338}.
\newblock


\bibitem[Li et~al\mbox{.}(2023)]%
        {frozen_ecg}
\bibfield{author}{\bibinfo{person}{Jun Li}, \bibinfo{person}{Che Liu}, \bibinfo{person}{Sibo Cheng}, \bibinfo{person}{Rossella Arcucci}, {and} \bibinfo{person}{Shenda Hong}.} \bibinfo{year}{2023}\natexlab{}.
\newblock \showarticletitle{Frozen Language Model Helps ECG Zero-Shot Learning}.
\newblock \bibinfo{journal}{\emph{arXiv preprint arXiv:2303.12311}} (\bibinfo{year}{2023}).
\newblock


\bibitem[Liu et~al\mbox{.}(2018)]%
        {cnn_ts3}
\bibfield{author}{\bibinfo{person}{Chien-Liang Liu}, \bibinfo{person}{Wen-Hoar Hsaio}, {and} \bibinfo{person}{Yao-Chung Tu}.} \bibinfo{year}{2018}\natexlab{}.
\newblock \showarticletitle{Time series classification with multivariate convolutional neural network}.
\newblock \bibinfo{journal}{\emph{IEEE Transactions on industrial electronics}} \bibinfo{volume}{66}, \bibinfo{number}{6} (\bibinfo{year}{2018}), \bibinfo{pages}{4788--4797}.
\newblock


\bibitem[Makowski et~al\mbox{.}(2021)]%
        {Makowski2021}
\bibfield{author}{\bibinfo{person}{Dominique Makowski}, \bibinfo{person}{Tam Pham}, \bibinfo{person}{Zen~J. Lau}, \bibinfo{person}{Jan~C. Brammer}, \bibinfo{person}{François Lespinasse}, \bibinfo{person}{Hung Pham}, \bibinfo{person}{Christopher Schölzel}, {and} \bibinfo{person}{S.~H.~Annabel Chen}.} \bibinfo{year}{2021}\natexlab{}.
\newblock \showarticletitle{NeuroKit2: A Python toolbox for neurophysiological signal processing}.
\newblock \bibinfo{journal}{\emph{Behavior Research Methods}} (\bibinfo{year}{2021}), \bibinfo{pages}{1--8}.
\newblock


\bibitem[Min et~al\mbox{.}(2021)]%
        {llm_survey1}
\bibfield{author}{\bibinfo{person}{Bonan Min}, \bibinfo{person}{Hayley Ross}, \bibinfo{person}{Elior Sulem}, \bibinfo{person}{Amir Pouran~Ben Veyseh}, \bibinfo{person}{Thien~Huu Nguyen}, \bibinfo{person}{Oscar Sainz}, \bibinfo{person}{Eneko Agirre}, \bibinfo{person}{Ilana Heinz}, {and} \bibinfo{person}{Dan Roth}.} \bibinfo{year}{2021}\natexlab{}.
\newblock \bibinfo{title}{Recent Advances in Natural Language Processing via Large Pre-Trained Language Models: A Survey}.
\newblock
\newblock
\showeprint[arxiv]{2111.01243}~[cs.CL]


\bibitem[Motin et~al\mbox{.}(2017)]%
        {Motin2017}
\bibfield{author}{\bibinfo{person}{Mohammod~Abdul Motin}, \bibinfo{person}{Chandan~Kumar Karmakar}, {and} \bibinfo{person}{Marimuthu Palaniswami}.} \bibinfo{year}{2017}\natexlab{}.
\newblock \showarticletitle{Ensemble empirical mode decomposition with principal component analysis: A novel approach for extracting respiratory rate and heart rate from photoplethysmographic signal}.
\newblock \bibinfo{journal}{\emph{IEEE Journal of Biomedical and Health Informatics}} \bibinfo{volume}{22}, \bibinfo{number}{3} (\bibinfo{year}{2017}), \bibinfo{pages}{766--774}.
\newblock


\bibitem[Nilsson et~al\mbox{.}(2000)]%
        {Nilsson2000}
\bibfield{author}{\bibinfo{person}{Lena Nilsson}, \bibinfo{person}{Anders Johansson}, {and} \bibinfo{person}{Sigga Kalman}.} \bibinfo{year}{2000}\natexlab{}.
\newblock \showarticletitle{Monitoring of respiratory rate in postoperative care using a new photoplethysmographic technique}.
\newblock \bibinfo{journal}{\emph{Journal of Clinical Monitoring and Computing}}  \bibinfo{volume}{16} (\bibinfo{year}{2000}), \bibinfo{pages}{309--315}.
\newblock


\bibitem[Pereira et~al\mbox{.}(2019)]%
        {pereira2019supervised}
\bibfield{author}{\bibinfo{person}{Tania Pereira}, \bibinfo{person}{Kais Gadhoumi}, \bibinfo{person}{Mitchell Ma}, \bibinfo{person}{Xiuyun Liu}, \bibinfo{person}{Ran Xiao}, \bibinfo{person}{Rene~A Colorado}, \bibinfo{person}{Kevin~J Keenan}, \bibinfo{person}{Karl Meisel}, {and} \bibinfo{person}{Xiao Hu}.} \bibinfo{year}{2019}\natexlab{}.
\newblock \showarticletitle{A supervised approach to robust photoplethysmography quality assessment}.
\newblock \bibinfo{journal}{\emph{IEEE journal of biomedical and health informatics}} \bibinfo{volume}{24}, \bibinfo{number}{3} (\bibinfo{year}{2019}), \bibinfo{pages}{649--657}.
\newblock


\bibitem[Pimentel et~al\mbox{.}(2016)]%
        {Pimentel2016}
\bibfield{author}{\bibinfo{person}{Marco A.~F. Pimentel}, \bibinfo{person}{Alistair E.~W. Johnson}, \bibinfo{person}{Peter~H. Charlton}, \bibinfo{person}{Drew Birrenkott}, \bibinfo{person}{Peter~J. Watkinson}, \bibinfo{person}{Lionel Tarassenko}, {and} \bibinfo{person}{David~A. Clifton}.} \bibinfo{year}{2016}\natexlab{}.
\newblock \showarticletitle{Toward a robust estimation of respiratory rate from pulse oximeters}.
\newblock \bibinfo{journal}{\emph{IEEE Transactions on Biomedical Engineering}} \bibinfo{volume}{64}, \bibinfo{number}{8} (\bibinfo{year}{2016}), \bibinfo{pages}{1914--1923}.
\newblock


\bibitem[Reiss et~al\mbox{.}(2019)]%
        {Reiss2019}
\bibfield{author}{\bibinfo{person}{Attila Reiss}, \bibinfo{person}{Ina Indlekofer}, \bibinfo{person}{Philip Schmidt}, {and} \bibinfo{person}{Kristof~Van Laerhoven}.} \bibinfo{year}{2019}\natexlab{}.
\newblock \showarticletitle{Deep PPG: Large-scale heart rate estimation with convolutional neural networks}.
\newblock \bibinfo{journal}{\emph{Sensors}} \bibinfo{volume}{19}, \bibinfo{number}{14} (\bibinfo{year}{2019}), \bibinfo{pages}{3079}.
\newblock


\bibitem[Schmidt et~al\mbox{.}(2018)]%
        {Schmidt2018}
\bibfield{author}{\bibinfo{person}{Philip Schmidt}, \bibinfo{person}{Attila Reiss}, \bibinfo{person}{Robert Duerichen}, \bibinfo{person}{Claus Marberger}, {and} \bibinfo{person}{Kristof~Van Laerhoven}.} \bibinfo{year}{2018}\natexlab{}.
\newblock \showarticletitle{Introducing WESAD, a multimodal dataset for Wearable Stress and Affect Detection}. In \bibinfo{booktitle}{\emph{ICMI}}. \bibinfo{address}{Boulder, USA}.
\newblock


\bibitem[Tang et~al\mbox{.}(2020)]%
        {cnn_ts1}
\bibfield{author}{\bibinfo{person}{Wensi Tang}, \bibinfo{person}{Guodong Long}, \bibinfo{person}{Lu Liu}, \bibinfo{person}{Tianyi Zhou}, \bibinfo{person}{Jing Jiang}, {and} \bibinfo{person}{Michael Blumenstein}.} \bibinfo{year}{2020}\natexlab{}.
\newblock \showarticletitle{Rethinking 1d-cnn for time series classification: A stronger baseline}.
\newblock \bibinfo{journal}{\emph{arXiv preprint arXiv:2002.10061}} (\bibinfo{year}{2020}), \bibinfo{pages}{1--7}.
\newblock


\bibitem[Torres-Soto and Ashley(2020)]%
        {TorresSoto2020}
\bibfield{author}{\bibinfo{person}{Jessica Torres-Soto} {and} \bibinfo{person}{Euan~A. Ashley}.} \bibinfo{year}{2020}\natexlab{}.
\newblock \showarticletitle{Multi-task deep learning for cardiac rhythm detection in wearable devices}.
\newblock \bibinfo{journal}{\emph{NPJ Digital Medicine}}  \bibinfo{volume}{3} (\bibinfo{year}{2020}), \bibinfo{pages}{116}.
\newblock
\urldef\tempurl%
\url{https://doi.org/10.1038/s41746-020-00320-4}
\showDOI{\tempurl}


\bibitem[Vaswani et~al\mbox{.}(2023)]%
        {transformer}
\bibfield{author}{\bibinfo{person}{Ashish Vaswani}, \bibinfo{person}{Noam Shazeer}, \bibinfo{person}{Niki Parmar}, \bibinfo{person}{Jakob Uszkoreit}, \bibinfo{person}{Llion Jones}, \bibinfo{person}{Aidan~N. Gomez}, \bibinfo{person}{Lukasz Kaiser}, {and} \bibinfo{person}{Illia Polosukhin}.} \bibinfo{year}{2023}\natexlab{}.
\newblock \bibinfo{title}{Attention Is All You Need}.
\newblock
\newblock
\showeprint[arxiv]{1706.03762}~[cs.CL]


\bibitem[Voisin et~al\mbox{.}(2019)]%
        {Shen2019}
\bibfield{author}{\bibinfo{person}{Maxime Voisin}, \bibinfo{person}{Yichen Shen}, \bibinfo{person}{Alireza Aliamiri}, \bibinfo{person}{Anand Avati}, \bibinfo{person}{Awni Hannun}, {and} \bibinfo{person}{Andrew Ng}.} \bibinfo{year}{2019}\natexlab{}.
\newblock \showarticletitle{Ambulatory atrial fibrillation monitoring using wearable photoplethysmography with deep learning}. In \bibinfo{booktitle}{\emph{Proceedings of the 25th ACM SIGKDD International Conference on Knowledge Discovery \& Data Mining}}. \bibinfo{pages}{1909--1916}.
\newblock


\bibitem[Wang et~al\mbox{.}(2023a)]%
        {Wang2023EMBC}
\bibfield{author}{\bibinfo{person}{Weinan Wang}, \bibinfo{person}{Pedram Mohseni}, \bibinfo{person}{Kevin~L Kilgore}, {and} \bibinfo{person}{Laleh Najafizadeh}.} \bibinfo{year}{2023}\natexlab{a}.
\newblock \showarticletitle{Demographic Information Fusion Using Attentive Pooling In CNN-GRU Model For Systolic Blood Pressure Estimation}. In \bibinfo{booktitle}{\emph{Annual International Conference of the IEEE Engineering in Medicine \& Biology Society (EMBC)}}. \bibinfo{pages}{1--4}.
\newblock


\bibitem[Wang et~al\mbox{.}(2023b)]%
        {Wang2023}
\bibfield{author}{\bibinfo{person}{Weinan Wang}, \bibinfo{person}{Pedram Mohseni}, \bibinfo{person}{Kevin~L. Kilgore}, {and} \bibinfo{person}{Laleh Najafizadeh}.} \bibinfo{year}{2023}\natexlab{b}.
\newblock \showarticletitle{PulseDB: A large, cleaned dataset based on MIMIC-III and VitalDB for benchmarking cuff-less blood pressure estimation methods}.
\newblock \bibinfo{journal}{\emph{Frontiers in Digital Health}}  \bibinfo{volume}{4} (\bibinfo{year}{2023}), \bibinfo{pages}{1090854}.
\newblock


\bibitem[Wang et~al\mbox{.}(2021a)]%
        {Wang2021CurrLearning}
\bibfield{author}{\bibinfo{person}{Xin Wang}, \bibinfo{person}{Yudong Chen}, {and} \bibinfo{person}{Wenwu Zhu}.} \bibinfo{year}{2021}\natexlab{a}.
\newblock \showarticletitle{A survey on curriculum learning}.
\newblock \bibinfo{journal}{\emph{IEEE Transactions on Pattern Analysis and Machine Intelligence}} \bibinfo{volume}{44}, \bibinfo{number}{9} (\bibinfo{year}{2021}), \bibinfo{pages}{4555--4576}.
\newblock


\bibitem[Wang et~al\mbox{.}(2021b)]%
        {Wang2021}
\bibfield{author}{\bibinfo{person}{Yingfan Wang}, \bibinfo{person}{Haiyang Huang}, \bibinfo{person}{Cynthia Rudin}, {and} \bibinfo{person}{Yaron Shaposhnik}.} \bibinfo{year}{2021}\natexlab{b}.
\newblock \showarticletitle{Understanding how dimension reduction tools work: an empirical approach to deciphering t-SNE, UMAP, TriMAP, and PaCMAP for data visualization}.
\newblock \bibinfo{journal}{\emph{The Journal of Machine Learning Research}} \bibinfo{volume}{22}, \bibinfo{number}{1} (\bibinfo{year}{2021}), \bibinfo{pages}{9129--9201}.
\newblock


\bibitem[Woo et~al\mbox{.}(2022)]%
        {Woo2022}
\bibfield{author}{\bibinfo{person}{Gerald Woo}, \bibinfo{person}{Chenghao Liu}, \bibinfo{person}{Doyen Sahoo}, \bibinfo{person}{Akshat Kumar}, {and} \bibinfo{person}{Steven Hoi}.} \bibinfo{year}{2022}\natexlab{}.
\newblock \bibinfo{title}{CoST: Contrastive learning of disentangled seasonal-trend representations for time series forecasting}.
\newblock \bibinfo{howpublished}{arXiv preprint arXiv:2202.01575}.
\newblock


\bibitem[Yang et~al\mbox{.}(2023)]%
        {biot}
\bibfield{author}{\bibinfo{person}{Chaoqi Yang}, \bibinfo{person}{M~Brandon Westover}, {and} \bibinfo{person}{Jimeng Sun}.} \bibinfo{year}{2023}\natexlab{}.
\newblock \showarticletitle{BIOT: Biosignal Transformer for Cross-data Learning in the Wild}. In \bibinfo{booktitle}{\emph{Thirty-seventh Conference on Neural Information Processing Systems}}.
\newblock


\bibitem[Yue et~al\mbox{.}(2022)]%
        {Yue2022}
\bibfield{author}{\bibinfo{person}{Zhihan Yue}, \bibinfo{person}{Yujing Wang}, \bibinfo{person}{Juanyong Duan}, \bibinfo{person}{Tianmeng Yang}, \bibinfo{person}{Congrui Huang}, \bibinfo{person}{Yunhai Tong}, {and} \bibinfo{person}{Bixiong Xu}.} \bibinfo{year}{2022}\natexlab{}.
\newblock \showarticletitle{Ts2vec: Towards universal representation of time series}. In \bibinfo{booktitle}{\emph{Proceedings of the AAAI Conference on Artificial Intelligence}}, Vol.~\bibinfo{volume}{36}. \bibinfo{pages}{8980--8987}.
\newblock


\bibitem[Zhang et~al\mbox{.}(2014)]%
        {Zhang2014}
\bibfield{author}{\bibinfo{person}{Zhilin Zhang}, \bibinfo{person}{Zhouyue Pi}, {and} \bibinfo{person}{Benyuan Liu}.} \bibinfo{year}{2014}\natexlab{}.
\newblock \showarticletitle{TROIKA: A general framework for heart rate monitoring using wrist-type photoplethysmographic signals during intensive physical exercise}.
\newblock \bibinfo{journal}{\emph{IEEE Transactions on Biomedical Engineering}} \bibinfo{volume}{62}, \bibinfo{number}{2} (\bibinfo{year}{2014}), \bibinfo{pages}{522--531}.
\newblock


\bibitem[Zhao et~al\mbox{.}(2017)]%
        {cnn_ts2}
\bibfield{author}{\bibinfo{person}{Bendong Zhao}, \bibinfo{person}{Huanzhang Lu}, \bibinfo{person}{Shangfeng Chen}, \bibinfo{person}{Junliang Liu}, {and} \bibinfo{person}{Dongya Wu}.} \bibinfo{year}{2017}\natexlab{}.
\newblock \showarticletitle{Convolutional neural networks for time series classification}.
\newblock \bibinfo{journal}{\emph{Journal of Systems Engineering and Electronics}} \bibinfo{volume}{28}, \bibinfo{number}{1} (\bibinfo{year}{2017}), \bibinfo{pages}{162--169}.
\newblock


\bibitem[Zhao et~al\mbox{.}(2023)]%
        {llm_survey2}
\bibfield{author}{\bibinfo{person}{Wayne~Xin Zhao}, \bibinfo{person}{Kun Zhou}, \bibinfo{person}{Junyi Li}, \bibinfo{person}{Tianyi Tang}, \bibinfo{person}{Xiaolei Wang}, \bibinfo{person}{Yupeng Hou}, \bibinfo{person}{Yingqian Min}, \bibinfo{person}{Beichen Zhang}, \bibinfo{person}{Junjie Zhang}, \bibinfo{person}{Zican Dong}, {et~al\mbox{.}}} \bibinfo{year}{2023}\natexlab{}.
\newblock \showarticletitle{A survey of large language models}.
\newblock \bibinfo{journal}{\emph{arXiv preprint arXiv:2303.18223}} (\bibinfo{year}{2023}).
\newblock


\bibitem[Zheng et~al\mbox{.}(2023a)]%
        {Zheng2023}
\bibfield{author}{\bibinfo{person}{Xiaochen Zheng}, \bibinfo{person}{Xingyu Chen}, \bibinfo{person}{Manuel Schürch}, \bibinfo{person}{Amina Mollaysa}, \bibinfo{person}{Ahmed Allam}, {and} \bibinfo{person}{Michael Krauthammer}.} \bibinfo{year}{2023}\natexlab{a}.
\newblock \bibinfo{title}{SimTS: Rethinking Contrastive Representation Learning for Time Series Forecasting}.
\newblock \bibinfo{howpublished}{arXiv preprint arXiv:2303.18205}.
\newblock


\bibitem[Zheng et~al\mbox{.}(2023b)]%
        {Zheng2023UTransBPNet}
\bibfield{author}{\bibinfo{person}{Yali Zheng}, \bibinfo{person}{Qing Liu}, \bibinfo{person}{Jingyuan Hong}, \bibinfo{person}{Shenghao Wu}, {and} \bibinfo{person}{Yuanting Zhang}.} \bibinfo{year}{2023}\natexlab{b}.
\newblock \bibinfo{title}{UTransBPNet: A General Deep Learning Model for Cuffless Blood Pressure Estimation under Activities}.
\newblock \bibinfo{howpublished}{Authorea Preprints}.
\newblock


\bibitem[Zhou and Selvaraj(2020)]%
        {Zhou2020}
\bibfield{author}{\bibinfo{person}{Menglian Zhou} {and} \bibinfo{person}{Nandakumar Selvaraj}.} \bibinfo{year}{2020}\natexlab{}.
\newblock \showarticletitle{Heart rate monitoring using sparse spectral curve tracing}. In \bibinfo{booktitle}{\emph{Annual International Conference of the IEEE Engineering in Medicine \& Biology Society (EMBC)}}. IEEE, \bibinfo{pages}{5347--5352}.
\newblock


\bibitem[Zhu et~al\mbox{.}(2020)]%
        {Zhu2020}
\bibfield{author}{\bibinfo{person}{Zhenglin Zhu}, \bibinfo{person}{Yawang Wang}, \bibinfo{person}{Xichuan Zhou}, \bibinfo{person}{Liuqing Yang}, \bibinfo{person}{Geng Meng}, {and} \bibinfo{person}{Ze Zhang}.} \bibinfo{year}{2020}\natexlab{}.
\newblock \showarticletitle{SWAV: a web-based visualization browser for sliding window analysis}.
\newblock \bibinfo{journal}{\emph{Scientific Reports}} \bibinfo{volume}{10}, \bibinfo{number}{1} (\bibinfo{year}{2020}), \bibinfo{pages}{149}.
\newblock


\end{thebibliography}

\appendix

\end{document}